\documentclass{aa} 
\usepackage{graphicx}
\usepackage{txfonts}
\usepackage{xcolor}
\usepackage{multirow}
\usepackage{braket}
\usepackage{verbatim}
\usepackage{hyperref}
\usepackage{soul}

\newcommand{\todo}[1]{{\color{red}[TODO: #1]}}
\newcommand{\lm}[1]{{\color{blue}[LM: #1]}}
\newcommand{\md}[1]{{\color{purple}[MD: #1]}}

\begin{document}

  \title{Astrophysical constraints from future measurements of the kinetic Sunyaev-Zel'dovich power spectrum}



  \author{L. McBride\thanks{elizabeth.mc-bride@universite-paris-saclay.fr}
     \inst{1}
     \and
     A. Gorce\inst{1}
     \and
     M. Douspis\inst{1,2}
     \and
     R. Meriot\inst{2,3}
     \and
     B. Semelin\inst{2}
     \and
     L.~T.~Hergt\inst{4}
     \and
     S. Ilić\inst{4}
     \and
     M. Muñoz-Echeverría\inst{5}
    \and
     E. Pointecouteau\inst{5}
     \and
     L. Salvati\inst{1}
     \and
     M. Tristram\inst{4}
     }

  \institute{Université Paris-Saclay, CNRS, Institut d’Astrophysique Spatiale, 91405 Orsay, France
    \and
    Observatoire de Paris, PSL Research University, Sorbonne Université, CNRS, LUX, 75014 Paris, France
    \and
    Imperial College London, Blackett Laboratory, Prince Consort Road, London SW7 2AZ, UK
    \and
    IJCLab, Université Paris-Saclay, CNRS/IN2P3, IJCLab, 91405 Orsay, France
    \and
    IRAP, CNRS, Université de Toulouse, CNES, UT3-UPS, Toulouse, France
    }

  \date{Received ***; accepted ***}

 
 \abstract
  {High-precision measurements of the Cosmic Microwave Background (CMB) will soon allow for the unprecedented detection of small-scale secondary anisotropies, such as the kinetic Sunyaev-Zel’dovich (kSZ) effect.}
  {Linking the kSZ power spectrum to the properties of ionising sources would provide an opportunity to use such observations to access astrophysical and cosmological information from the Epoch of Reionisation, including the morphology of ionised regions, while simultaneously improving CMB analyses. The aim of this work is to assess this potential of the kSZ power spectrum to measure reionisation-era galaxy properties.}
  {We repurpose the publicly available \textsc{LoReLi~II} simulations, which track the evolution of neutral hydrogen during reionisation, to generate a training set of patchy kSZ angular power spectra. We then train an emulator using neural network regression in order to allow for efficient Bayesian inference,  and conduct forecasts assuming mock observations from current and future CMB experiments.}
 {We find that measurements of the kSZ power spectrum from such surveys can provide meaningful constraints on several of the astrophysical model parameters of the \textsc{LoReLi~II} suite, including the ionising escape fraction for which we expect a 14\% relative error, on average. They also provide an independent measurement of the CMB optical depth, marginalised over the astrophysics and with error bars competitive with the cosmic variance limit from large scale surveys. 
}
  {The kSZ power spectrum offers a promising avenue for probing the properties of reionisation-era galaxies and providing an independent measurement of the CMB optical depth with upcoming CMB experiments. Since the error budget of our mock observations is dominated by emulator reconstruction errors, we expect our results could be further improved with a more extended simulation training set.}


  \maketitle
%
\section{Introduction}

Relic photons from the surface of last scattering, known collectively as the Cosmic Microwave Background (CMB), form an unparalleled snapshot of the infant Universe and have provided some of the best constraints on cosmological parameters to date \citep{Mather, Fixsen, PlanckCosmo18, Planck_params, BATMAMmatthieu}. However these photons are not immune to their passage through the evolving Universe, but instead have a probability to undergo scattering events since their decoupling at Recombination. These interactions imprint subtle distortions, or secondary anisotropies, atop the primary signal, and carry valuable information about both the growth of large-scale structure (LSS), and the process of reionisation, by which the Universe transforms from neutral hydrogen gas to its ionised state as seen today \citep{Aghanim_2008}. By teasing apart these different signals, it is possible to not only further isolate the primary CMB, thereby improving cosmological constraints, but also extract information about astrophysics and the evolution of the Universe from what was historically a contaminant \citep{Douspis2006, Douspis_2022_I,GorceDouspis_2022}.

CMB distortions arise from several broad classes of interactions. One such class is known as the Sunyaev--Zel’dovich effect (SZ), in which CMB photons have a chance to scatter off of free electrons as they traverse the LSS of the Universe. During this process they can be Doppler boosted via inverse Compton scattering if those free electrons have a bulk motion with respect to their local environment. CMB photons then emerge from this long journey through the ionising intergalactic medium (IGM) with a cumulative energy shift built up from each interaction, in a process known as the kinetic SZ effect \citep[kSZ, ][]{kSZ1, kSZ2}. In this way, the kSZ distortion links cosmological observations to the astrophysics that drive galaxy formation. The total kSZ signal is often decomposed into two regimes, the patchy (pkSZ) contribution, which arises during the Epoch of Reionisation (EOR) due to ionised regions, or bubbles, and the homogenous component (hkSZ) which is dominated by the late-time (or post-reionisation) ambient free electron background. 

In standard CMB power spectrum analyses, the presence of the kSZ effect has often been accounted for with templates \citep[e.g.][]{Planck2018_Likelihood, Camphuis_SPT_2025, louis_atacama_2025} that have little impact on cosmological constraints \citep{BATMAMmatthieu} but which remove the possibility of extracting the rich information embedded in the SZ distortions \citep{Douspis_2022_I, GorceDouspis_2022}. However in recent years, gains in precision from advances in both instrumentation and data analysis have made disentangling these anisotropies feasible, and first constraints on the amplitude of the kSZ power spectrum by \cite{kSZmeasurement1}, \cite{choi2020}, and \cite{kSZmeasurement2} have increased interest in gaining a greater understanding of the signal, which is a considerable challenge. 
 The kSZ signal is an integrated quantity along the line of sight and thus one must coherently model the history of the Universe from the surface of last scattering around $z=1100$ to the end of reionisation (for the patchy contribution). Additionally, the patchy kSZ signal is highly dependent on the physics of reionisation, a notoriously complex and non-linear process. It is this very dependency that makes the kSZ signal an exquisite probe of the EOR, with the potential to constrain both the timing \citep{McQuinn_2006, Iliev_2008,  Mesinger_2012} and duration \citep{Zahn2012_reio, Planckreio_2016} of reionisation, as well the optical depth to reionisation, $\tau_{\mathrm{CMB}}$ \citep{Battagliatau2016, morefrenemies}. Reionisation also depends heavily on the ionising sources themselves, and more recent works have focused on relating their astrophysical properties to the kSZ signal. 
Since the EOR is difficult to describe analytically, the kSZ signal is often studied numerically, for example using hydrodynamical simulations \citep{Dolag2016}, or via ray tracing through a simulated light cone \citep{Battaglia_2010, Mesinger_2012, Battaglia_2013, Chen_AMBER}.  
An alternate approach consists in analytically deriving the kSZ power spectrum from simulation products, such as the power spectrum of ionisation \citep{ourfrenemies} or electron density \citep{GorceIlic_2020} fluctuations. Using simulations based on \textsc{emma} \citep{Aubert_2015}, an adaptive mesh refinement code designed to simulate the reionisation epoch, \cite{GorceIlic_2020} derived the kSZ power spectrum through a parameterisation of the electron overdensity power spectrum. 
However, this model was calibrated on a limited dataset: only six simulations, all run with the same astrophysical parameters. Thus, the authors were unable to investigate the impact of astrophysics on the signal. The recently available \textsc{LoReLi~II} database \citep{MeriotSemelin_2025} comprises three orders of magnitude more simulations, generated by varying five astrophysical parameters. These parameters affect the ionising properties of galaxies, thereby altering the reionisation timeline and its morphology, both of which leave their imprint on the kSZ signal. 

With the \textsc{LoReLi~II} dataset as our starting point, in this work we investigate the impact of ionising source properties on the kSZ signal. In Section~\ref{sec:data}, we introduce the \textsc{LoReLi~II} database. In Section~\ref{sec:methods}, we describe how we build a dataset of patchy kSZ angular power spectra from the \textsc{LoReLi~II} simulations, which we use to train an emulator. With this emulator as our forward model, we forecast the constraining power of mock kSZ observations on astrophysical parameters in Section\,\ref{sec:results}. We discuss our results in Section\,\ref{sec:discuss} and summarise our findings in Section\,\ref{sec:fin}. 

\section{Data} \label{sec:data}


In order to study the impact of astrophysical parameters on the patchy kSZ signal, we build our analysis from the \textsc{LoReLi~II} simulations \citep{MeriotSemelin_2024, MeriotSemelin_2025}. However, these simulations were designed to study the evolution of neutral hydrogen in the intergalactic medium (IGM) via the high-redshift 21\,cm signal, and as such they do not explicitly output the kSZ signal. Instead, it is possible to reconstruct the kSZ angular power spectrum by exploiting its dependence on the power spectrum of electron overdensity fluctuations,  $P_{\mathrm{ee}}(k, z)$, and the ionisation history, $x_e(z)$, reconstructed from the \textsc{LoReLi~II} simulation cubes (see \cite{GorceIlic_2020} and Section\,\ref{subsec:Gorce}). Below, we provide an overview of the \textsc{LoReLi~II} simulations and discuss the scope of the dataset.

\begin{figure}
\centering
\includegraphics[width=.8\columnwidth]{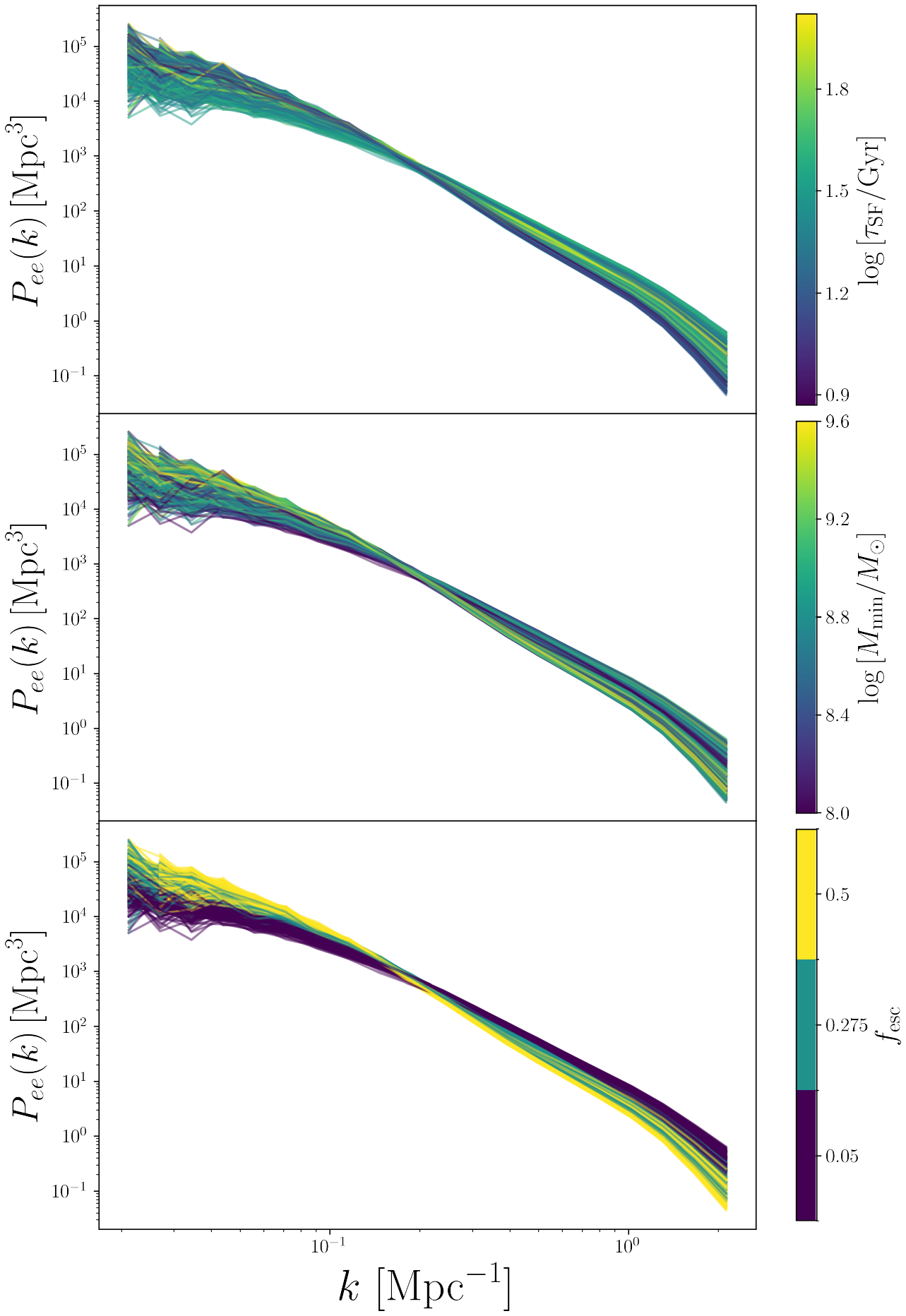}
\caption{Electron overdensity power spectra derived from the \textsc{LoReLi~II} database shown at the midpoint of reionisation, $x_\mathrm{HII}(z) = 0.5$. The spectra are coloured by the value of the three astrophysical parameters varied in our analysis.}
\label{fig:LoReLi}
\end{figure}

\subsection{\textsc{LoReLi~II}} \label{subsec:LoReLi}
The \textsc{LoReLi~II} database is a collection of about ten thousand 3D radiative hydrodynamical cosmological simulation cubes designed to capture the transition of the Universe from neutral to nearly fully ionised.
The simulations are generated with the \textsc{Licorice} N-body simulation code \citep{semelin2007lyman, baek2009simulated, semelin201721ssd, semelin2023accurate, MeriotSemelin_2024} and contain both baryons and dark matter particles. Radiative transfer effects for the two ionising bands, UV and X-ray, are also included using Monte-Carlo ray-tracing. All simulations are run in a cosmological volume of $(200 \,h^{-1}\text{Mpc})^3$ with a grid size of $256^3$. For each simulation run, a density box is initialised at $z =53.48$ with an independent realisation of the primordial density field and then evolved forward. The first snapshot is saved at a redshift of $z=22.0$, with subsequent snapshots saved at intermittent timestamps that depend on the specific simulation. A Planck 2018 cosmology \citep{PlanckCosmo18} is assumed for all simulations. 
 While originally conceived for the study of the cosmological 21\,cm brightness temperature, snapshots of the simulated matter density and ionisation fields are also saved, which allows the \textsc{LoReLi~II} simulations to be repurposed for a broader range of science, including an investigation of the patchy kSZ signal. 
In this work we show the wide utility of such a simulation set, and the ability to extract additional science even for which the design is suboptimal. In Sec.~\ref{sec:discuss} we list some avenues for improvement for future simulations seeking to maximise their use cases for EOR studies.

\subsection{Astrophysical Parameters} \label{subsec:params}
We provide a brief summary of the astrophysical parameters varied in the \textsc{LoReLi~II} suite, along with their range of input values. 
Each set of parameter values is run only once, along with a unique realisation of the initial conditions. While this allows the \textsc{LoReLi~II} database to cover an expansive range in parameter space, it also means we are unable to mitigate the effects of large-scale sample variance on our primary derived data product, the electron overdensity power spectrum, $P_{ee}$.


\vspace{3pt}
\textbf{X-ray Production Efficiency ($f_X$)}. The X-ray production efficiency governs the X-ray luminosity of galaxies, following \cite{furlanetto2006cosmology}:
\begin{align}
L_X = 3.4 \times10^{40} f_X \left(\frac{SFR}{1 M_{\odot} \mathrm{yr}^{-1}} \right) \text{ergs}^{-1}.
\end{align}
The X-ray production efficiency is sampled at 13 logarithmically spaced values on [0.1, 10].

\vspace{3pt}
\textbf{Hard X-ray fraction ($r_{\mathrm{H}/\mathrm{S}}$)}.
The simulations contain two X-ray emission mechanisms that contribute to the total flux: X-ray binaries with energetic spectra (that is, typical photon energy greater than 2~keV), and main sequence stars (typical photon energy less than 2~keV). The hard X-ray fraction is defined as the ratio of the two contributions: $r_{\mathrm{H}/\mathrm{S}} \equiv f_X^{XRB} /f_X$, which sets the overall hardness of the total X-ray flux. Setting the value of $r_{\mathrm{H}/\mathrm{S}}$ also has the secondary effect of changing the UV flux, as the two photon flux regimes share a pooled energy budget within the simulation. In the \textsc{LoReLi~II} suite, there are six linearly spaced values of the hard X-ray fraction on [0, 1].

\vspace{3pt}
\textbf{Gas Conversion Timescale ($\tau_\mathrm{SF}$)}.
The gas conversion timescale governs how efficiently galaxies turn gas into stars, thus determining the rate of star formation in galaxies. The sampling in parameter space covers the range $\tau_{\mathrm{SF}} \in [7\,\mathrm{Gyr}, 105\,\mathrm{Gyr}]$, but is coupled with the minimum halo mass.

\vspace{3pt}
\textbf{Minimum Halo Mass ($M_{\text{min}}$)}.
Haloes must reach some critical mass, defined as the minimum halo mass, $M_{\text{min}}$, before their star formation is sufficient to produce enough ionising radiation to form bubbles. This threshold has a large impact on their subsequent morphology, as a larger $M_{\text{min}}$ results in larger and rarer bubbles. The sampled values range over $M_{\text{min}} \in [10^8 \text{M}_{\odot}, 4\times10^9 \text{M}_{\odot}]$, covaried with the gas conversion timescale.

In the \textsc{LoReLi~II} database, only certain pairs $(\tau_{\text{SF}}$, $M_{\text{min}})$ are sampled. As they both have a direct and inversely related effect on star formation rate densities, the values of these two parameters are always covaried in order to produce star formation rate densities consistent with current constraints from observations from $5 \leq z \leq 10$ \citep{McLeod, Oesch, Bouwens_2021}. While yielding reasonable histories within this redshift window, each pair can yield vastly different star formation scenarios at higher redshifts (see the discussion in \citealt{MeriotSemelin_2024}). We further discuss the impact of the coupling of these two parameters on our analysis in Sect.\,\ref{sec:methods} and Appendix\,\ref{appendix:2dprior}.

\vspace{3pt}
\textbf{Ionising Escape Fraction ($f_{\text{esc}}$)}.
Not all ionising photons produced by galaxies reach the IGM,  as they can be absorbed by the dense gas contained within haloes. The portion that does escape is set by the ionising escape fraction, $f_{\text{esc}}$. The \textsc{LoReLi~II} simulation code defines two escape fractions: a pre- and a post-threshold value, where the threshold criteria is the density of the neutral hydrogen surrounding a source. In the dataset, only the post-threshold escape fraction is varied and we refer to this parameter simply as the ionising escape fraction. The sampled values are $f_{\text{esc}} \in \{0.05, 0.275, 0.5 \}$. 

In Figures~\ref{fig:LoReLi} (with additional ionisation fractions plotted in Appendix\,\ref{appendix:astroparams}) and \ref{fig:xe} we show the electron overdensity power spectra and ionisation histories, respectively, for the entire dataset coloured as a function three of the five \textsc{LoReLi} parameters: the gas conversion timescale, the minimum halo mass, and the ionising escape fraction. The parameter with the clearest impact on the electron power spectrum is the ionising escape fraction. The trend for the ionisation histories is similar, and we elect to only show the dependence on $f_{\text{esc}}$ in Figure~\ref{fig:xe}. Intuitively, large values of the ionising escape fraction mean a large number of available ionising photons, which abets a speedy reionisation process. Note that, while important for the 21\,cm brightness temperature, we expect the kSZ signal to be only weakly sensitive to the two X-ray parameters, as X-ray photons are too energetic to efficiently ionise neutral hydrogen. We discuss this assumption and its implications in Appendix\,\ref{appendix:Xrays}.

\begin{figure}
\centering
\includegraphics[width=0.8\columnwidth]{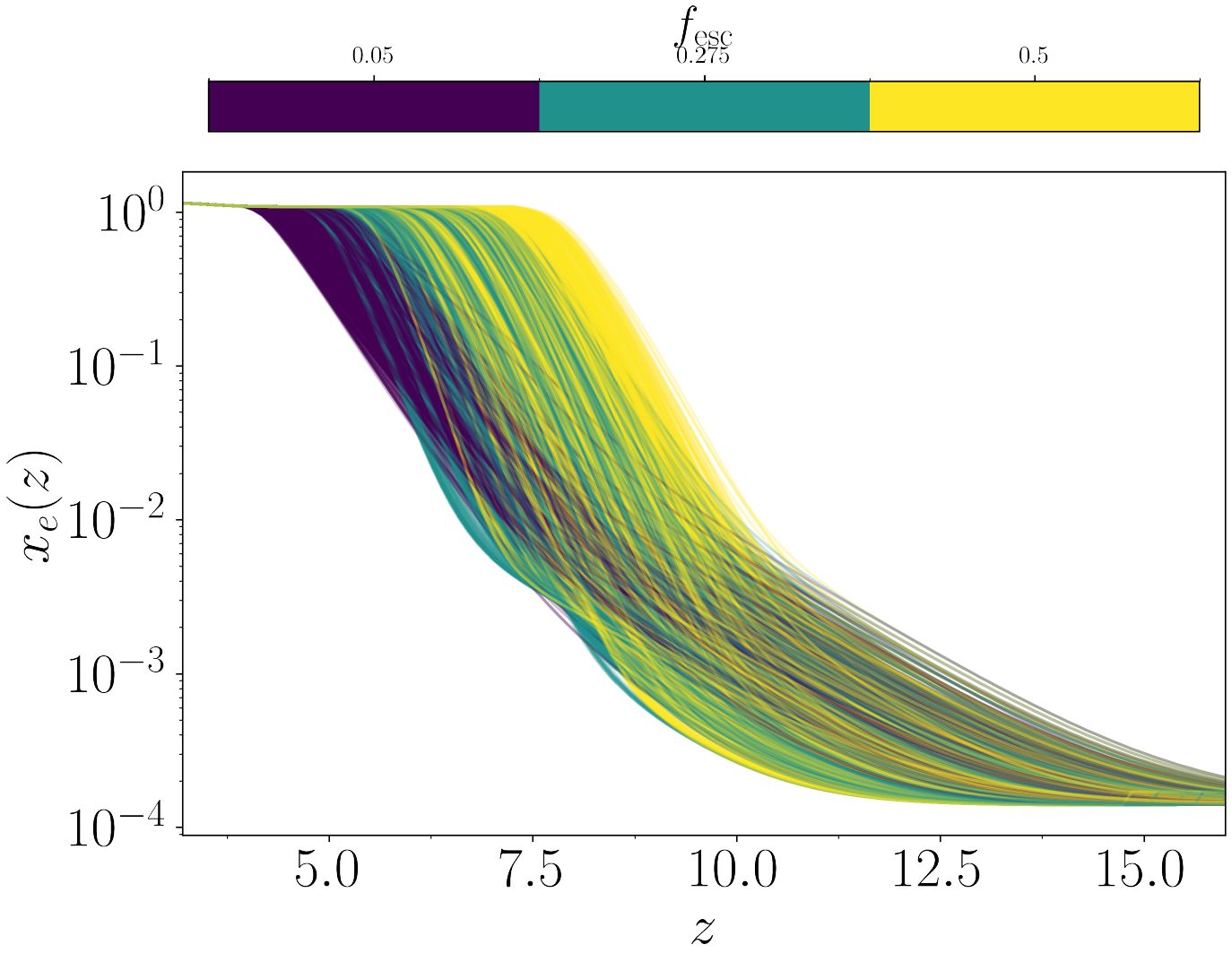}
\caption{Ionisation histories $x_e(z)$, defined as the volume-averaged ionisation fraction, for all simulations used in our analysis, coloured by the value of $f_\mathrm{esc}$, the parameter with the greatest observed impact on the kSZ power spectrum. 
}
\label{fig:xe}
\end{figure}



\section{Methods}\label{sec:methods}
In this section, we lay out the methodology to build our kSZ catalogue and conduct our forecasts. We introduce the analytic formalism for the kSZ signal in Section~\ref{subsec:Gorce} and detail the reconstruction procedure --- starting from \textsc{LoReLi~II} data cubes through to the patchy kSZ angular power spectrum --- in Section~\ref{subsec:kSZfromLoReLi}. We then describe our emulator training and check its accuracy across the parameter space in Section~\ref{subsec:emulator}. In Sec.~\ref{subsec:framework}, we describe the Bayesian analysis framework we apply to mock datasets for two idealised CMB survey configurations.



\subsection{Deriving the kSZ power spectrum} \label{subsec:Gorce}  
Our kSZ power spectrum reconstruction is based on the prescription described in \cite{GorceIlic_2020}, which relates the angular kSZ power spectrum, $C_\ell^{\mathrm{kSZ}}$, to the electron overdensity power spectrum, $P_{\mathrm{ee}}(k,z)$. The basic intuition is that the kSZ signal arises from interactions between CMB photons and free electrons whose distribution, both spatially and in redshift, will directly impact the kSZ signal. We formalise this intuition here.
The temperature anisotropies due to the kSZ effect are given by
\begin{equation} \label{eq:deltaT_kSZ}
    \delta T_{\text{kSZ}}(\hat{\mathbf{n}}) = \frac{\sigma_T}{c} \int \frac{d\eta}{dz} \frac{dz}{1+z} e^{-\tau_\mathrm{CMB}(z)}\, n_e(z)\, \mathbf{v} \cdot \hat{\mathbf{{n}}}, 
\end{equation}
and depend on the Thomson scattering cross-section, $\sigma_T$, the speed of light, $c$, the comoving distance $\eta$ to a redshift $z$, the line of sight component of the electron peculiar velocity, $\mathbf{v} \cdot \hat{\mathbf{n}}$, and the Thomson optical depth, 
\begin{equation} \label{eq:tau_CMB}
\tau_{\mathrm{CMB}}(z) = c \sigma_T \int_0^{z} \frac{\bar{n}_e(z')}{H(z')}(1+z')^2 dz',
\end{equation}
which itself depends on the Hubble parameter, $H(z)$.
Both $\delta T_{\text{kSZ}}$ and $\tau_{\mathrm{CMB}}$ are sensitive to the free electron number density,
\begin{equation} \label{eq:n_e}
n_e(\hat{\mathbf{n}},z)=\bar{n}_e(z)\left[1+\delta_e(\hat{\mathbf{n}},z)\right],
\end{equation}
where $\bar{n}_e(z)\equiv x_e(z) n_b(z)$ is the free electron density at redshift $z$, averaged over spatial fluctuations, for $x_e$ the ionisation history, defined as the volume-averaged fraction of free electrons normalised to the total number of baryons, $n_b(z)$. Finally, $\delta_e$ is the free electron overdensity field, whose power spectrum is $P_{ee}(k,z)$. 
The scale-dependent variance information of the kSZ-induced temperature fluctuations can then be quantified via its power spectrum as a function of multipole $\ell$,
\begin{equation} \label{eq:Cell}
C_{\ell}^{\mathrm{kSZ}} = \frac{8\pi^2}{(2\ell+1)^3} \frac{\sigma_T^2}{c^2} \int \frac{\bar{n}_e^2(z)}{(1+z)^2} \Delta_{B,e}^2(\ell/\eta,z) e^{-2\tau(z)} \eta \frac{d\eta}{dz} dz,
\end{equation} 
where $\Delta_{B,e}^2(k, z) = k^3 P_{B,e}(k,z) / 2\pi^2$. The power spectrum of the curl component of the momentum is given by
\begin{equation} \label{eq:PBe}
\begin{aligned}
&P_{B,e}(k, z) = f^2(z) \dot{a}^2(z) \int d^3k' \,(1-\mu^2)\, \times \\
& \quad \left[ \frac{1}{k'^2} P_{ee}(|\mathbf{k}-\mathbf{k}'|) P_{\delta\delta}^\mathrm{lin}(k',z) - \frac{ b_{\delta e}(k',z)}{|\mathbf{k} -\mathbf{k}'|^2}\, b_{\delta e}(|\mathbf{k}-\mathbf{k}'|, z)\right], 
\end{aligned}
\end{equation}
which depends on the scale factor, $a$, the linear growth factor, $f$, as well as the linear total matter power spectrum, $P_{\delta \delta}^{\mathrm{lin}}$, and the bias, $b_{\delta e}$, defined as the ratio of the free electron overdensity power spectrum over the non-linear matter power spectrum, and is integrated over $\mu=\hat{\mathbf{k}} \cdot \hat{\mathbf{k}}'$.


\subsection{Reconstruction from simulation cubes} \label{subsec:kSZfromLoReLi}

To form our kSZ catalogue, we begin by extracting $P_{ee}(k,z)$ and $x_e(z)$ from each \textsc{LoReLi~II} simulation in the database, which is easily done with the ionisation fraction and density cubes available at each redshift snapshot. However, due to the nature of the dataset, some analysis choices must be made. 
Because only one realisation of each field is available, the reconstructed electron overdensity power spectra are strongly affected by sample variance. This is particularly true on larger scales, which contribute most strongly to the spherical harmonic modes 
where the patchy kSZ signal is at its maximum. In order to mitigate unphysical structure due to this sample variance, we choose to bin coarsely in Fourier mode k, using $n=10$ evenly spaced logarithmic bins, which was found by examination to provide a balance between reducing spurious peaks without washing out all pertinent features.

First, in order to investigate the patchy contribution to the kSZ, we integrate Eq.~\eqref{eq:deltaT_kSZ} up to the end of reionisation. However, the simulations in the \textsc{LoReLi~II} database terminate at varying ionisation fractions, with the majority ending between $.9 < x_\text{HII} < .985$ volume-averaged hydrogen ionisation fraction. In order to  homogenise the endpoint of reionisation across all simulations, while maximising the number of reconstructed kSZ spectra, we choose to integrate only up to the redshift (typically $4.8 \lesssim z \lesssim 8$) which corresponds to 97\% hydrogen ionisation. This leads to a reduction in the overall amplitude of the spectrum, but is enforced consistently across all simulations in order to standardise the amount of signal lost across the catalogue. This truncation means the kSZ signal is slightly underestimated, increasing the conservativeness of our forecasts. Additionally, we remove from the dataset simulations that do not reach 97\% ionisation. Because the lowest value of the ionising escape fraction ($f_{\mathrm{esc}} = 0.05$) tends to lead to late ionisation (and thus an increased chance of early termination of the simulation run), we also disproportionately lose simulations with the lowest escape fraction. In total, we retain $n_{\text{sims}}=6\,796$ simulations to build our kSZ database.
Second, because of the finite size of the simulation, the largest scales cannot be accessed and there is a potential loss of kSZ power at low-$k$. We show in Appendix\,\ref{appendix:cuts} that, on average, the missing power from large scales and the $x_e<0.97$ cut lead to a $\lesssim5\%$ signal loss.

Finally, helium reionisation is not accounted for in the \textsc{LoReLi~II} simulations. To correct for this, we assume simultaneous first reionisation of helium and hydrogen \citep[see][]{Furlanetto_2008, McQuinn_2009}, and simply scale the derived ionisation histories by the correct electron fraction when including \ion{He}{I}, $f_e = f_\ion{H}{I} + f_\ion{He}{I} = 1.08$. 
In order to compare with current measurements of the optical depth we assume a simple $\tanh$ model \citep{Lewis_2008} for the second reonisation of helium, centred at $z=3.5$. 



\subsection{Emulator training} \label{subsec:emulator}
With the method described above, we reconstruct the angular kSZ spectra at the points in parameter space covered by the \textsc{LoReLi~II} dataset. However, to assess the constraining power of a kSZ measurement on astrophysical parameters, we need a forward model to predict the observed kSZ at \emph{any} point in the parameter space. To achieve this, we train an ensemble of neural networks to predict the patchy kSZ angular power spectrum given a set of astrophysical parameters. While we use the emulator to interpolate between parameter values, we do not attempt to extrapolate and predict the spectrum for values outside the range covered in the \textsc{LoReLi~II} database (see Sec.~\ref{subsec:params}). We train our data on all five astrophysical parameters as it was found to improve the emulator performance.

The regression is performed using a neural network (NN) implemented with the \texttt{keras} package \citep{keras}. We observe that different random seeds can introduce variability in the predicted signal, since the performance of any individually trained emulator is sensitive to both random weight initialisation and data partitioning. To overcome this issue, we adopt bootstrap aggregating (bagging), wherein an ensemble of emulators is trained on different subsets of the data, and their outputs are combined to produce the final emulated signal. Each emulator is trained on 80\% of the total dataset, while the remaining 20\% is retained as a pristine validation set that is not used by any emulator within the ensemble during training. 

Each NN takes in the training data, normalised by subtracting the mean and dividing by the standard deviation in order to reduce the dynamic range. During training, the network evaluates the quality of the weights using a standard mean squared error (MSE) loss function\footnote{We find that custom loss functions, designed to take into account multipole-dependent errors on the kSZ spectrum, do not yield improved performance.}. Our architecture comprises four layers and is implemented via the \texttt{keras} \texttt{Sequential} API. It begins with an input layer, followed by two fully connected (dense) hidden layers, each containing 32 neurons with \texttt{LeakyReLU} activation. The final output layer uses a linear activation to generate continuous predictions. Training is run for 500 epochs (but with early stopping allowed) using the Adam optimiser and a \texttt{ReduceLROnPlateau} learning rate scheduler that reduces the learning rate when validation loss stops improving.
Given the modest size and sparsity of our dataset, overfitting is a concern, and several aspects of our regression setup are designed to mitigate it. The low number of neurons and limited training epochs reduce the model's capacity to overfit the training data. Additionally, L2 regularisation is applied to each dense layer to penalise large weights and encourage generalisation. The regularisation strength is set to $\lambda = 0.001$ for the input layer and is gradually decreased in deeper layers to allow increased flexibility after the initial learning phase. Each regression is also initialised with a specific, but unique, random seed in order to improve the reproducibility of our final emulator.

\begin{figure}
\centering
\includegraphics[width=0.8\columnwidth]{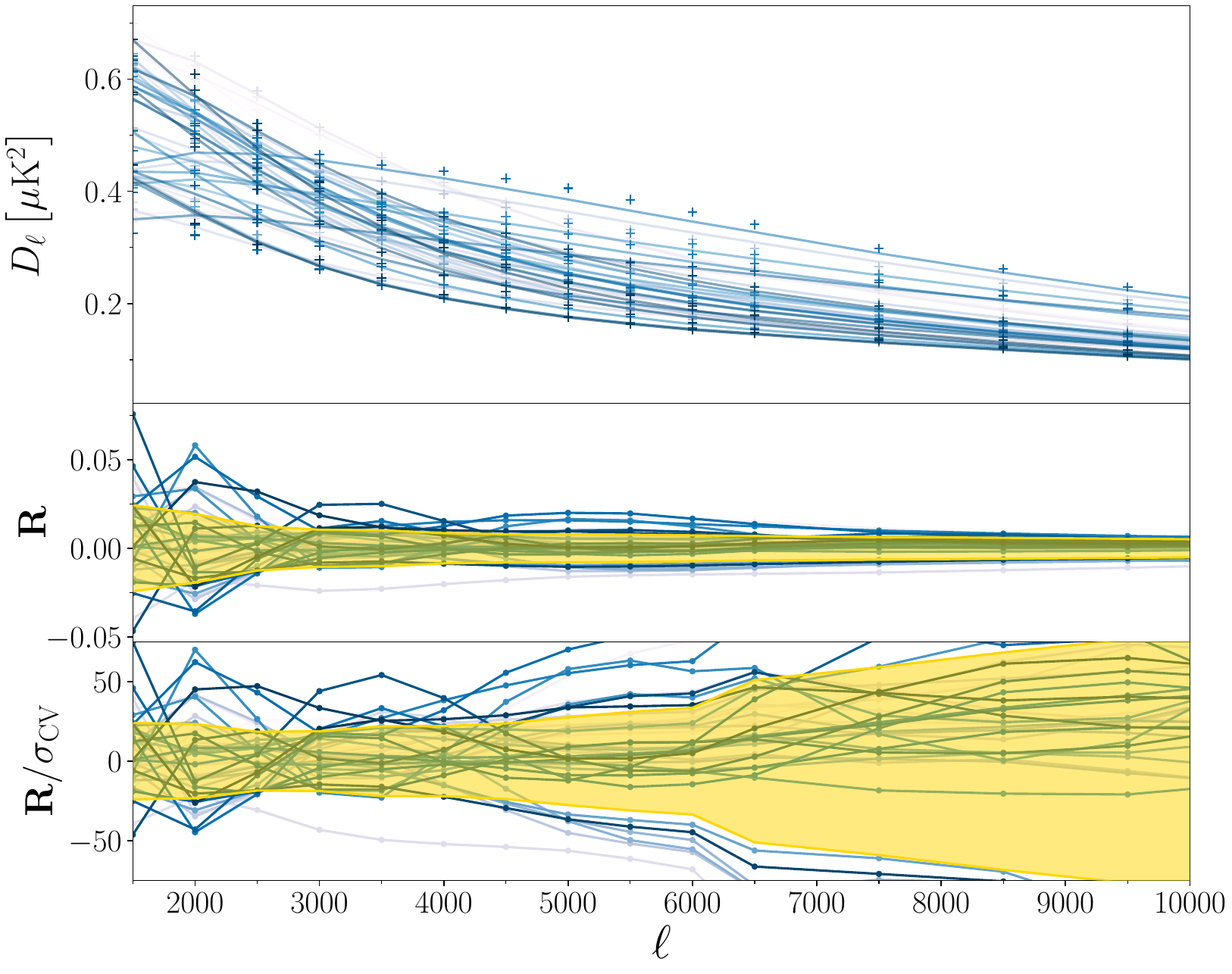}
\caption{Emulated pkSZ spectra from the validation dataset, overlaid with the true values (marked with $+$) of each spectrum (top panel). The residual of the averaged emulator outputs (predicted $-$ true) is shown both unnormalised (middle panel) and normalised (bottom panel) by the cosmic variance of our fiducial kSZ signal. The estimated emulator $1\sigma$ error, as given by Eq.\,\eqref{eq:emu_err} overlaid in yellow.}
\label{fig:kemu}
\end{figure}

In addition to stability, another key concern is the accuracy of the regression model across the entire parameter space. A comparison between the emulated and true kSZ spectra is shown in Fig.\,\ref{fig:kemu}. Because of the sample variance which affects $P_{ee}$, the lowest multipoles of the reconstructed kSZ spectra are highly variable, leading all neural network iterations to consistently underperform on that range. To mitigate this, we remove $\ell < 1500$ data points from the kSZ catalogue, and emulate fewer multipoles for $\ell > 6500$, which meaningfully improves the accuracy of the model. Overall, we emulate the kSZ power spectrum on the range $1500 \leq \ell \leq 6500$ with a $\Delta \ell= 500$ spacing, and increase to $\Delta \ell = 1000$ up to $\ell = 9500$.
We quantify the error in the emulated signals by using the reserved 20\% validation set and calculating the residual, $\mathbf{R} = y_\mathrm{emu} - y_\mathrm{true}$, defined as the difference between the true (input) kSZ spectrum reconstructed from a \textsc{LoReLi~II} simulation, and the prediction outputted by the emulator. We then take the standard deviation of this residual as the 1$\sigma$ error on our emulated signals,
\begin{align} \label{eq:emu_err}
   \sigma^2_{\mathrm{emu}}(\ell) = \Big\langle  \mathrm{Var}^{N_\mathrm{emu}}\left[R(\theta, \ell) \right] \Big\rangle_{N_\theta}, 
\end{align}
where the variance of the residual $R$ is found across emulators for a single set of parameter values $\theta$ before averaging over all  $N_\theta = 1359$ parameter sets in the validation simulations.


We present emulator accuracy results in Fig.~\ref{fig:kemu}. As can be seen in the middle panel, the residual is typically small (68\% of the validation spectra are recovered within $< 0.05\,\mu\mathrm{K}^2$) and larger at low $\ell$. Note that the residual is often larger than the cosmic variance of the patchy kSZ signal (bottom panel), which could lead to biased inference. 
Additionally, the average ratio between the true and emulated spectra is nearly unity at all multipoles, however for individual simulations there can be larger errors. 
As an example, for our fiducial simulation (Sec.~\ref{subsec:case}), the residual at $\ell=3000$ is roughly $\sim1.85$ larger than the standard deviation of the residuals at this multipole, $\sigma_{\ell=3000}=0.011$. While the average ratio $\braket{y_{\text{emu}} / y_{\text{true}}}$ is marginally greater than unity (implying that, in general, the emulator tends to overestimate the amplitude of the signal), for our fiducial simulation the signal is underestimated ($\braket{y_{\text{emu}} / y_{\text{true}}} = 0.96$). We note that emulator error is the leading source of uncertainty at most multipoles (see Fig.\,\ref{fig:surveys}), therefore in Sec.~\ref{sec:methods}, we introduce a nuisance parameter in order to mitigate this bias. Our emulator accuracy across the parameter space will impact our analysis, and we investigate its consequences in Sec.~\ref{subsec:stats}.

Finally, we train an emulator of the ionisation history which allows us to compute the reionisation optical depth, $\tau_{\mathrm{CMB}}$, as a function of the astrophysical parameters\footnote{Since we are interested in the contribution from reionisation, we integrate Eq.\,\eqref{eq:tau_CMB} from $z=20$ to the present day \citep{BATMAN!}.}, 
built with a similar architecture. For this second emulator, we achieve good accuracy, with a mean absolute error (MAE) $\sim 10^{-3}$ on the predicted optical depth.



\subsection{Forecast approach} \label{subsec:framework}

\subsubsection{Mock dataset} \label{subsec:surveys}
\begin{figure}
\centering
\includegraphics[width=\columnwidth]{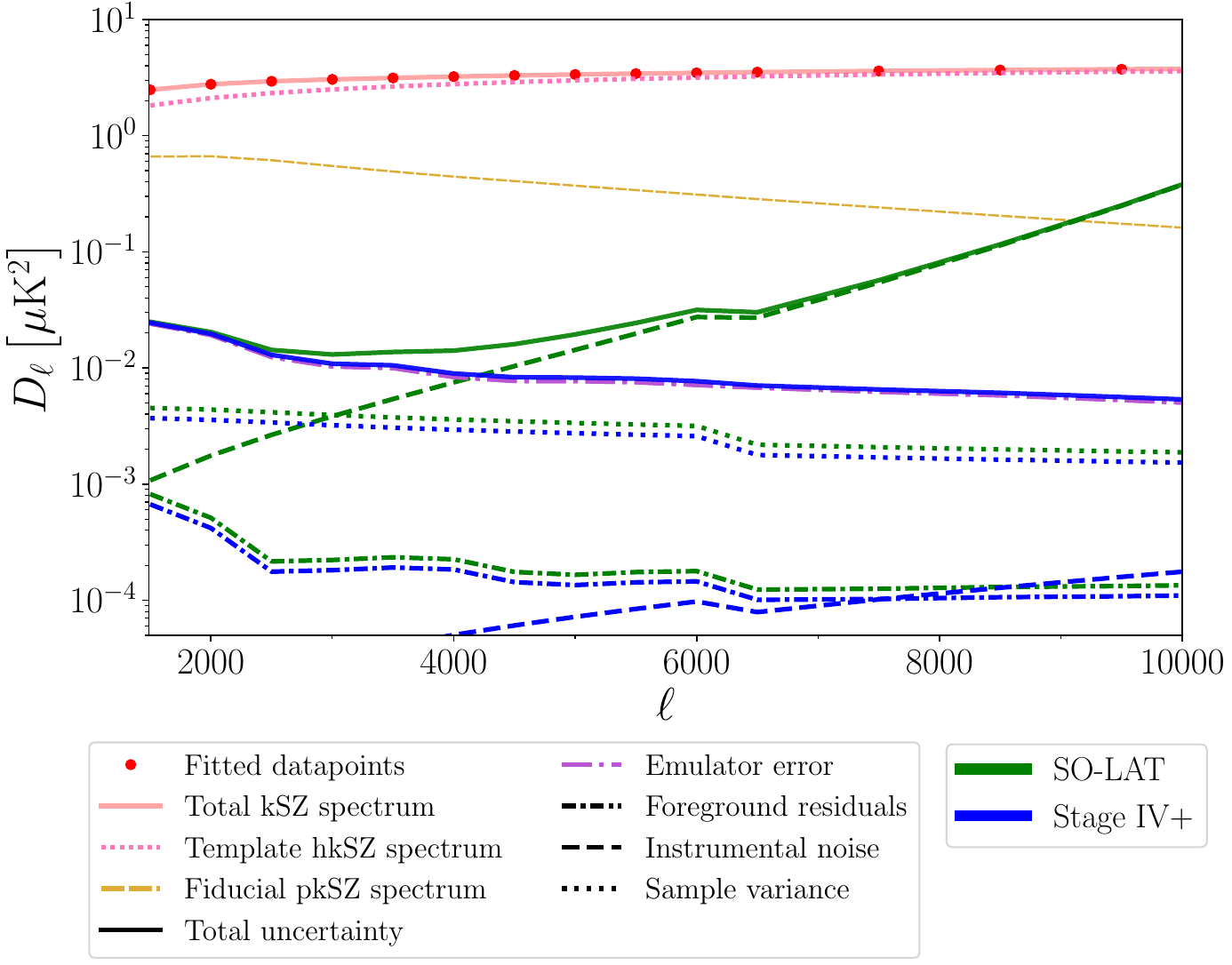}
\caption{Mock data points (red) and observational uncertainties for the two survey configurations considered in this work. Each individual contribution, consisting of sample variance (loose dotted), instrumental noise (loose dashed), and foreground residuals (dense dash-dotted), to the total assumed uncertainty (solid line) are shown for each survey. We also show the emulator error (violet long-dash-dotted line) -- identical between surveys. Both the patchy (gold dense dashed) and the homogeneous spectra (pink dense dotted) contribute to the total kSZ signal (red solid).}
\label{fig:surveys}
\end{figure}

We construct the fiducial mock kSZ power spectrum used in the forecast as the sum of a patchy and a late-time contributions. Our mock observations are composed of only the kSZ spectra and no added noise realisation, 
allowing us to isolate the bias on our inference arising strictly from the analysis framework. The pkSZ contribution is reconstructed from the \textsc{LoReLi~II} simulation with $\log{f_X}=-1.72$,  $r_{\mathrm{H/S}}= 0$,  $\tau_{\mathrm{SF}}=10^{1.28}\,\mathrm{Gyr}$,  $M_\mathrm{min}= 10^{8.53}\,M_{\odot}$, and $f_\mathrm{esc}=0.275$, yielding  $\tau_\mathrm{CMB}=0.0553$. This parameter set $\mathbf{\theta}_{\text{true}}$ is selected such that $f_{\mathrm{esc}}$ does not lie on a prior boundary and $\tau_\mathrm{CMB}$ is compatible with \textit{Planck} observations. The resulting pkSZ spectrum peaks at $\ell=2000$ with a maximum amplitude of $D_{\ell=2000}^{\mathrm{pkSZ}}=0.664 \, \mathrm{\mu K^2}$, and 
$D_{\ell=3000}^{\mathrm{pkSZ}}= 0.548 \, \mathrm{\mu K^2}$. For the homogeneous term, we use a template shape obtained by integrating the analytic model in \cite{GorceIlic_2020} from the end of reionisation to $z=0$, using a bias description of the electron overdensity power spectrum \citep{Shaw_2012}. We fix the amplitude such that $D^{\mathrm{hkSZ}}_{\ell=3000}=2.9 \,\mu \mathrm{K}^2$, consistent with SPT and \textit{Planck} observations \citep{GorceDouspis_2022}.
The mock dataset is shown in red in Fig.\,\ref{fig:surveys}.

We include different sources of uncertainties in the mock dataset, as laid out in \citet{CMBerrs}, and both shown in Fig.~\ref{fig:surveys}: the sample variance due to the finite number of observable modes (dotted line), and experimental noise (dashed line). Then the observational uncertainty for a $C_\ell$ measurement\footnote{For our analysis, we will present results in $D_{\ell} = \ell(\ell + 1)C_{\ell} / 2\pi$.} over a bin of width $\Delta \ell$ centred on multipole $\ell$ is
\begin{align}
\label{eq:obs_error_cl}
    \Delta C_{\ell} = \sqrt{\frac{2}{2\ell + 1} \frac{1}{f_{\mathrm{sky}}}} \left[ C_{\ell} + N_{\ell} \right] \frac{1}{\sqrt{\Delta \ell}},
\end{align}
and depends on the sky fraction, $f_{\mathrm{sky}}$, of the survey. For $N_{\ell}$, we assume instrument noise, characterised by the temperature map noise $\sigma_0$ and the beam full width half maximum $\Theta_{\mathrm{FWHM}}$, along with an added contribution due to imperfect foreground subtraction,
\begin{align}
N_{\ell} = \frac{\sigma_0^2}{2} \exp{\left[ \frac{\ell (\ell + 1) \Theta^2_{\mathrm{FWHM}}}{8 \ln{2}} \right]} + N_{\ell}^{\mathrm{fg}},
\end{align}
where our estimate of the foreground residuals error $N_{\ell}^{\mathrm{fg}}$ (shown in a densely dash-dotted line in Fig.\,\ref{fig:surveys}) is derived from the \texttt{HDlike} framework \citep{hdlike}. 
We consider two instruments: a Stage\,III (current) and an idealised Stage IV+ CMB experiment\footnote{Indeed, we have noted that for the sensitivity expected from Stage~VI experiments, the error budget is dominated by the emulator uncertainty. Therefore, results will be equivalent for any subsequent generation of telescope.}. For Stage III, we assume experiment specifications analogue to the Large Aperture Telescope \citep[LAT, ][]{SO_LAT2021} of the Simons Observatory \citep[SO,][]{SO_science}. For Stage IV+, we consider a high resolution survey inspired by projects such as CMB-S4 \citep{CMB-S4, CMB-S4_science} and CMB-HD \citep{CMB-HD}. The survey properties we use are given in Table\,\ref{table:telescopes}, and are based on the specifications listed in \cite{hdlike}. Finally, we assume the amplitude of the sample variance term is proportional to a clean kSZ signal measurement, with no contribution from primary anisotropies, and thus represents a best case scenario. 

In Fig.\,\ref{fig:surveys}, we compare the projected uncertainties for both surveys, as well as the modelling error due to our emulator. 
In terms of observational uncertainty, the instrumental noise contribution at low multipoles is subdominant, but foreground contamination (particularly from Galactic dust emission) remains a stubborn systematic and impedes clean measurements for $\ell < 1000$. At higher multipoles ($\ell > 10\,000$), instrumental noise dominates over the potential signal. At intermediate multipoles however, there exists a window of opportunity for a clean measurement of the kSZ signal. Thus for both surveys considered, we fit the multipole range $\ell \in [1500,10000]$, with the exclusion of multipoles $1000 < \ell < 1500$ being due to the emulator training described in Sec.~\ref{subsec:emulator}.

\begin{table}
\caption{Technical specifications corresponding to the telescope configurations assumed when generating our mock datasets.} 
\label{table:telescopes} 
\centering 
\begin{tabular}{c c c c} 
\multirow{2}{4em}{Telescope} & Sky & FWHM & Noise  \\
 & fraction &(arcmin) &($\mu$K\,arcmin)\\
\hline 
SO-LAT & 0.4 & 1.5 & 6.0 \\
Stage IV+ & 0.6 & 0.42 & 0.7 \\
\hline 
\end{tabular}
\end{table}


\subsubsection{Posteriors} \label{subsec:Bayes}
Our strategy to determine parameter constraints is to map out the high probability regions of the posterior distribution given by Bayes theorem. We numerically approximate this posterior distribution by running a Markov Chain Monte Carlo (MCMC) sampler with the $\texttt{zeus}$ package \citep{zeus1, zeus2}, which traverses the parameter space with an Ensemble Slice Sampling method. The chains are run with $n_{\text{walkers}}=12$ for enough steps ($n_{\text{steps}}=10\,000$) to satisfy the Gelman-Rubin criteria for convergence $\hat{r} \sim 1$. 

As stated earlier, we fix $f_X$ and $r_{\mathrm{H/S}}$ to the true fiducial values and fit only for $\log{\tau_{\mathrm{SF}}}$, $\log{M_\mathrm{min}}$, and $f_\mathrm{esc}$. We additionally fit the nuisance parameter $A_{\mathrm{bias}}$ to account for emulator bias. Given the multipole dependence of $A_{\mathrm{bias}}$ (Fig.~\ref{fig:kemu}), we divide each emulated spectrum by a template made of the mean ratio between the true and emulated spectra in the validation set, $\bar{y}(\ell) = \braket{y_{\mathrm{emu}}(\ell)/y_{\mathrm{true}}{(\ell)}}$. We do not fit the spectral shape of the homogenous signal, but its amplitude at $\ell=3000$, $A_{\mathrm{hkSZ}}$. Overall, we fit for $\vec{\theta}_{\text{fit}} = (\log{\tau_{\mathrm{SF}}}, \log{M}_{\text{min}}, f_{\text{esc}}, A_{\mathrm{bias}}, A_{\mathrm{hkSZ}})$. We also derive the optical depth, $\tau_{\mathrm{CMB}}$, for each sampled parameter values, using our ionisation history emulator. 

\subsubsection{Likelihood}
 We assume a Gaussian likelihood and sample its natural logarithm, 
 \begin{align}
     \ln{P(\mathbf{d} | \vec{\theta})} \sim - \frac{1}{2} (\mathbf{d} - \mathbf{m} (\vec{\theta}))^T \Sigma^{-1} (\mathbf{d} - \mathbf{m}(\vec{\theta})),
 \end{align}
where $P$ depends on the data $\mathbf{d}$ and the model, $\mathbf{m(\theta)}$, a function of the fitted parameters, $\mathbf{\theta}$.
For the data covariance, $\Sigma$, we assume a diagonal matrix -- a reasonable approximation given the size of our multipole bins. Each element is such that $\Sigma^2 = \sigma_{\ell \ell'}^2 \delta_{\ell \ell'}$, with $\sigma_{\ell \ell'}^2 = \sigma_{\mathrm{obs}}^2 + \sigma^2_{\mathrm{emu}}$, with $\sigma_\mathrm{obs}$ the experimental uncertainty (Eq.~\ref{eq:obs_error_cl}), and $\sigma_{\mathrm{emu}}$ the emulator error. This total uncertainty is overlaid (solid line) for both surveys in Fig.\,\ref{fig:surveys}.

\subsubsection{Priors} \label{subsec:priors}

Machine learning algorithms are susceptible to wild extrapolations outside the confines of their training set. Therefore, we impose flat three-dimensional priors corresponding to the ranges of the astrophysical parameter values used in \textsc{LoReLi~II}. We account for the coupling of the $\log{\tau_{\mathrm{SF}}}$ and $M_{\text{min}}$ parameters with an additional two-dimensional prior to ensure the emulator stays within their covaried region. This coupled $\tau_{\mathrm{SF}}$/$M_{\text{min}}$ prior strongly impacts the posterior distribution, as it effectively carves out forbidden regions in parameter space and subsequently distorts any quantile-based statistic (see Appendix~\ref{appendix:2dprior} and Sec.~\ref{subsec:stats}). 
In order to reduce this effect, 
following \citet{millea2018cosmic, BATMAN!}, 
we perform prior flattening on our posterior samples and inverse weight the 1D marginalised posterior histograms by the 1D prior for each parameter.

For the nuisance parameter, $A_{\mathrm{bias}}$, we assume a Gaussian prior, centred on unity, with a standard deviation equal to the dispersion of this ratio. All parameter priors are summarised in Table~\ref{table:priors}.


\subsubsection{Summary Statistics}
While we present the full posterior distributions in Sec.~\ref{sec:results}, we also provide a forecasted measurement, and associated uncertainty, which we report as the mean of the posterior samples and a credible interval (CI), respectively. For our credible interval we need a representative width which quantifies the high probability region while taking into account the effect of the coupled $\tau_{\mathrm{SF}}$/$M_{\text{min}}$ prior.
 We observe that when fitting a parameter whose true value lies at the edge of the allowed region, the strict prior boundary 
can artificially compress the width of the histograms, leading to a spuriously small credible interval (see Appendix~\ref{appendix:2dprior}). In order to take this compression into account, we build a conservative 68\% credible interval  which we call $\mathrm{CI}_{68+}$.
We calculate the quantile-based $68\%$ credible interval\footnote{The width of the posterior bounded by the 16th and 84th percentiles.} from the samples of each parameter and split it around the mean value into two effective `half'-widths\footnote{Which would be identical for a symmetric probability distribution.}.
We double the larger of the two half-widths to obtain our (conservative) forecasted measurement uncertainty, $\mathrm{CI}_{68+}$.

Additionally, we observe some recovered parameter values to be biased with respect to the input. To quantify this effect, we use the normalised bias\footnote{Readers more comfortable with frequentist notation may think of $b^*=1$ as a Bayesian analogue of a $1\sigma$ bias.}, 
\begin{equation} \label{eq:normedbias}
    b^* \equiv \frac{\hat{\theta} - \theta_{\mathrm{true}}}{\mathrm{CI}_{68+}/2}.
\end{equation}
The closer $|b^*|$ is to zero, the more accurate the reconstruction. In the remainder of this paper, we will consider values above one as a significant bias.

\begin{table}
\caption{Prior ranges and types for MCMC runs. Note that $\log \tau_{\mathrm{SF}}$ follows a complex prior following the input settings of the simulations:  $P_{\tau_{\mathrm{SF}}} : 4.54 \le \log{\tau_{\mathrm{SF}}/\mathrm{Gyr}} - 0.38 \log{M_{\mathrm{min}}/M_{\odot}} \le 5.09$. The prior distribution is visualised in Fig.\,\ref{fig:forecasts}.}
\label{table:priors}
\centering
\begin{tabular}{c c p{3.5cm}} 
\hline\hline
Parameter & Prior Range & Prior type \\
\hline
$\log{\tau_{\mathrm{SF}}/\mathrm{Gyr}}$ & $0.90-2.02$ 
    & $P_{\tau_{\mathrm{SF}}} $ \\
$\log{M_{\mathrm{min}}/M_{\odot}}$ & $8.00-9.60$ & Uniform \\
$f_\mathrm{esc}$ & $0.05-0.50$ & Uniform\\
$A_\mathrm{bias}$ & & $\mathcal{N}(1.0, 0.0380)$\\
$A_\mathrm{hksZ}$ & $0.00-10.0$ & Uniform \\
\hline
\end{tabular}
\end{table}

\section{Results} \label{sec:results}
We now investigate what impact each astrophysical parameter has on the kSZ signal (Sec.~\ref{subsec:impact}), and, reciprocally, how well these parameters can be inferred from a measurement of the kSZ power spectrum (Sec.~\ref{subsec:forecast}), first with a case study forecast for a single mock measurement of the kSZ spectrum, and then with a statistical analysis of the full database.

\subsection{Impact of astrophysics on the patchy kSZ spectrum} \label{subsec:impact}
\begin{figure}
\centering
\includegraphics[width=0.9\columnwidth]{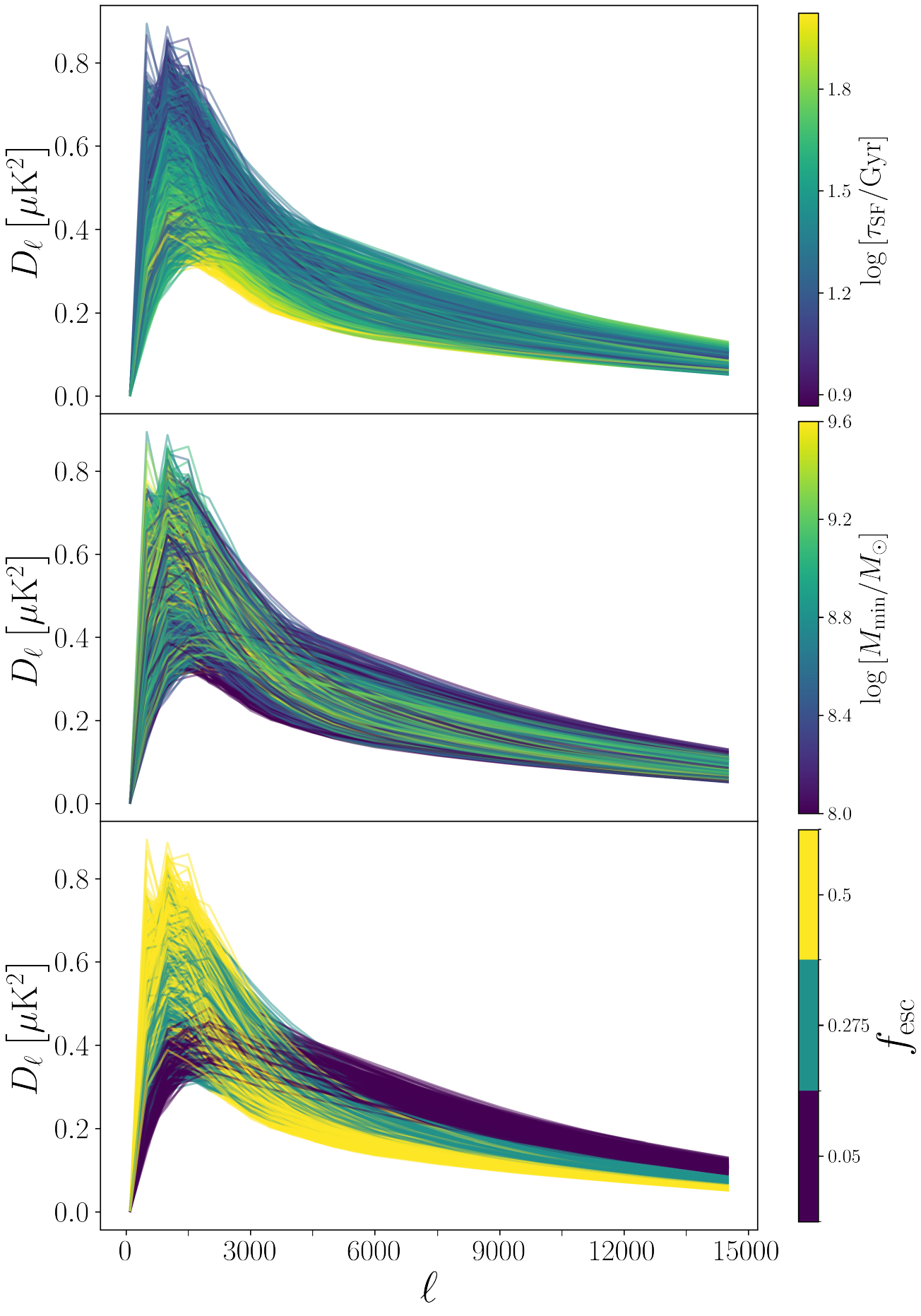}
\caption{The catalogue of patchy kSZ angular power spectra reconstructed from the \textsc{LoReLi~II} simulations, coloured by the value of each fitted astrophysical parameter.}
\label{fig:kSZ}
\end{figure}

\begin{figure}
\centering
\includegraphics[width=\columnwidth]{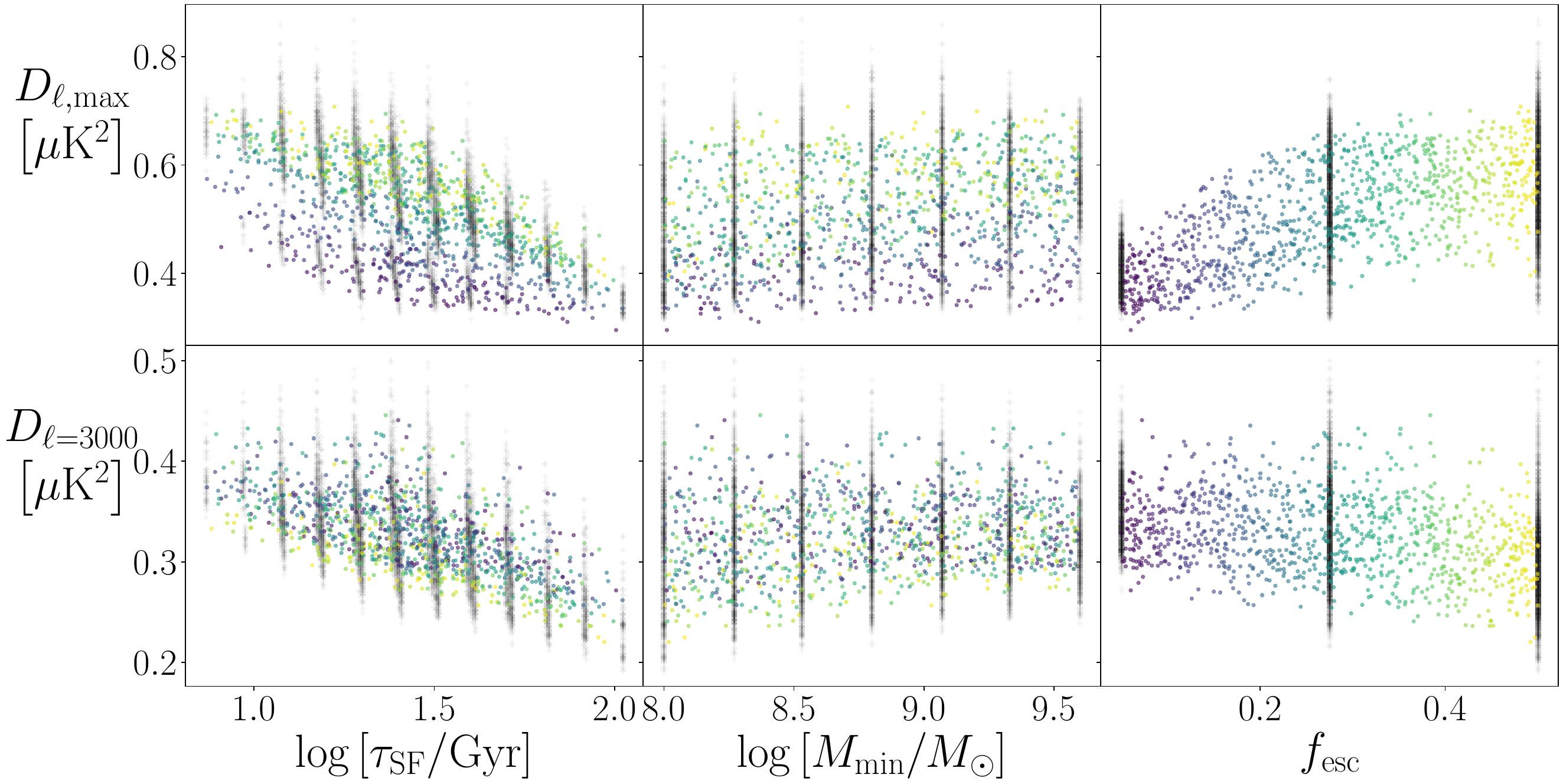}
\caption{Maximum amplitude of the patchy kSZ power spectrum (top row), and its amplitude at $\ell=3000$ (bottom row), as a function of the three fitted astrophysical parameters. For each row, the kSZ spectra are emulated from a uniform random draw of parameter values (coloured by the value of the ionising escape fraction from the draw) or taken from our training catalogue (black markers).}
\label{fig:trends}
\end{figure}

Figure~\ref{fig:kSZ} shows the patchy kSZ spectra reconstructed from the \textsc{LoReLi~II} dataset (Sec.~\ref{sec:data}), colour-coded by the values of the astrophysical parameters used as input to the simulations. Additionally, in Fig.~\ref{fig:trends}, we highlight two key features of the kSZ spectrum as a function of the input parameters: the maximum amplitude of the signal, and the amplitude at $\ell=3000$. These figures show that the patchy kSZ power spectrum is sensitive to the astrophysical parameters used to produce the \textsc{LoReLi~II} simulations, although to varying degrees. 

As already seen for the electron overdensity power spectra (Fig.~\ref{fig:LoReLi}), the value of the ionising escape fraction strongly impacts the shape and amplitude of the patchy kSZ spectrum. 
The ionising escape fraction sets how efficiently galaxies can inject ionising photons into the IGM, making it a key driver of not only the timing and duration of reionisation (Fig.\,\ref{fig:xe}), but also of the size of ionised bubbles, all of which are traced by the patchy kSZ power. Therefore, larger values of the escape fraction lead to earlier reionisation, pushing up the (maximum) amplitude of the power spectrum. Similarly, larger $f_\mathrm{esc}$ values lead to larger ionised bubbles \citep[e.g.,][]{SeilerHutter_2019}, such that the maximum patchy kSZ amplitude is reached at lower multipoles. 

The gas conversion timescale also has a large impact on the kSZ signal. A proxy for star formation, $\log{\tau_{\mathrm{SF}}}$ controls how efficiently galaxies can create the stars that are a major source of ionising flux, thus large values of $\log{\tau_{\mathrm{SF}}}$ impede reionisation by delaying star formation. This dampens the overall amplitude of the spectrum, as seen in Fig.~\ref{fig:trends}. Here, the impact on bubble morphology is less clear than for $f_\mathrm{esc}$. In Fig\,6 we see the additional influence of the ionising escape fraction on the maximum amplitude as a function of $\log{\tau_{\mathrm{SF}}}$ (top left), where it leads to a splitting of the values into two regimes.

The kSZ power spectrum exhibits sensitivity to the minimum halo mass, $M_{\text{min}}$, albeit more weakly than the other two parameters. The minimum halo mass is expected to have a large impact on the bubble morphology, and in turn on the shape of the spectrum, as observed for 21CMFAST simulations in \citet{GorceIlic_2020}. For example, a scenario where halos are required to grow significantly before they can reionise their local environment (large $M_\mathrm{min}$) will lead to fewer, but larger, bubbles. Conversely, if halos can ionise without being very massive, then reionisation will start early and consist of more, albeit smaller, bubbles. This bubble morphology affects the kSZ signal in two significant ways. The larger and rarer the bubbles, the larger the amplitude of the kSZ power spectrum. Additionally, it can shift the peak of the maximum amplitude, which is related to the characteristic bubble size. However, it is clear even by visual inspection that the impact of $M_{\text{min}}$ on the shape of the kSZ signal is not as strong as for the other two parameters. Roughly speaking the effects of $\log{\tau_{\mathrm{SF}}}$ and $M_{\text{min}}$ are inversely related, and this degeneracy (reinforced by their coupling in the parameter space) may wash out its impact. 

\begin{table*}
\caption{Forecasted parameter constraints and biases for a Stage\,III (SO-LAT-like) and Stage IV+ CMB survey. Best-fit values are the posterior means and uncertainties of the prior-flattened samples based on the conservative 68\% credible interval. The normalised bias, $b^*$ (Eq.~\ref{eq:normedbias}), is also shown.}
\label{tab:bestfits}
\renewcommand{\arraystretch}{1.3}
\centering
\begin{tabular}{l c 
                c c c  
                c c c  
               }
\hline\hline
Parameter & Fiducial
  & \multicolumn{2}{c}{SO-LAT-like} 
  & \multicolumn{2}{c}{Stage IV+} \\
 & value 
  & $\hat{\theta}\pm\mathrm{CI}_{68+}/2$ & $b^*$
  & $\hat{\theta}\pm\mathrm{CI}_{68+}/2$ & $b^*$ \\
\hline
$\log{\tau_{\mathrm{SF}}/\mathrm{Gyr}}$ & 1.28 
  & $1.18\pm 0.24$ & $-0.42$
  & $1.28 \pm 0.15$ & $-0.0010$ \\
$\log{M_{\mathrm{min}}/M_{\odot}}$ & 8.53 
  & $8.72\pm0.50$ & $0.38$ 
  & $8.48 \pm 0.31$ & $-0.16$ \\
$f_{\mathrm{esc}}$ & 0.275 
  & $0.33\pm0.072$ & $0.76$ 
  & $0.26 \pm .033$ & $-0.52$ \\
\hline
$A_{\mathrm{\mathrm{bias}}}$ & 1.040 
  & $1.01 \pm 0.035$ & $-1.01$ 
  & $1.06\pm0.024$ & $0.60$ \\
$A_{\mathrm{\mathrm{hkSZ}}}$ & 2.90 
  & $2.94\pm0.018$ & $2.0$ 
  & $2.90\pm0.0064$ & $-0.041$ \\
\hline\hline
$\tau_{\mathrm{CMB}}$ & 0.0553
  & $0.054\pm0.0023$ & $-0.43$ 
  & $0.053 \pm 0.0018$ & $-1.02$ \\
\hline
\end{tabular}
\end{table*}



\subsection{Forecasts} \label{subsec:forecast}

We now turn these qualitative observations into quantitative results through formal forecasts.

\begin{figure*}
\centering
\includegraphics[width=.75\textwidth]{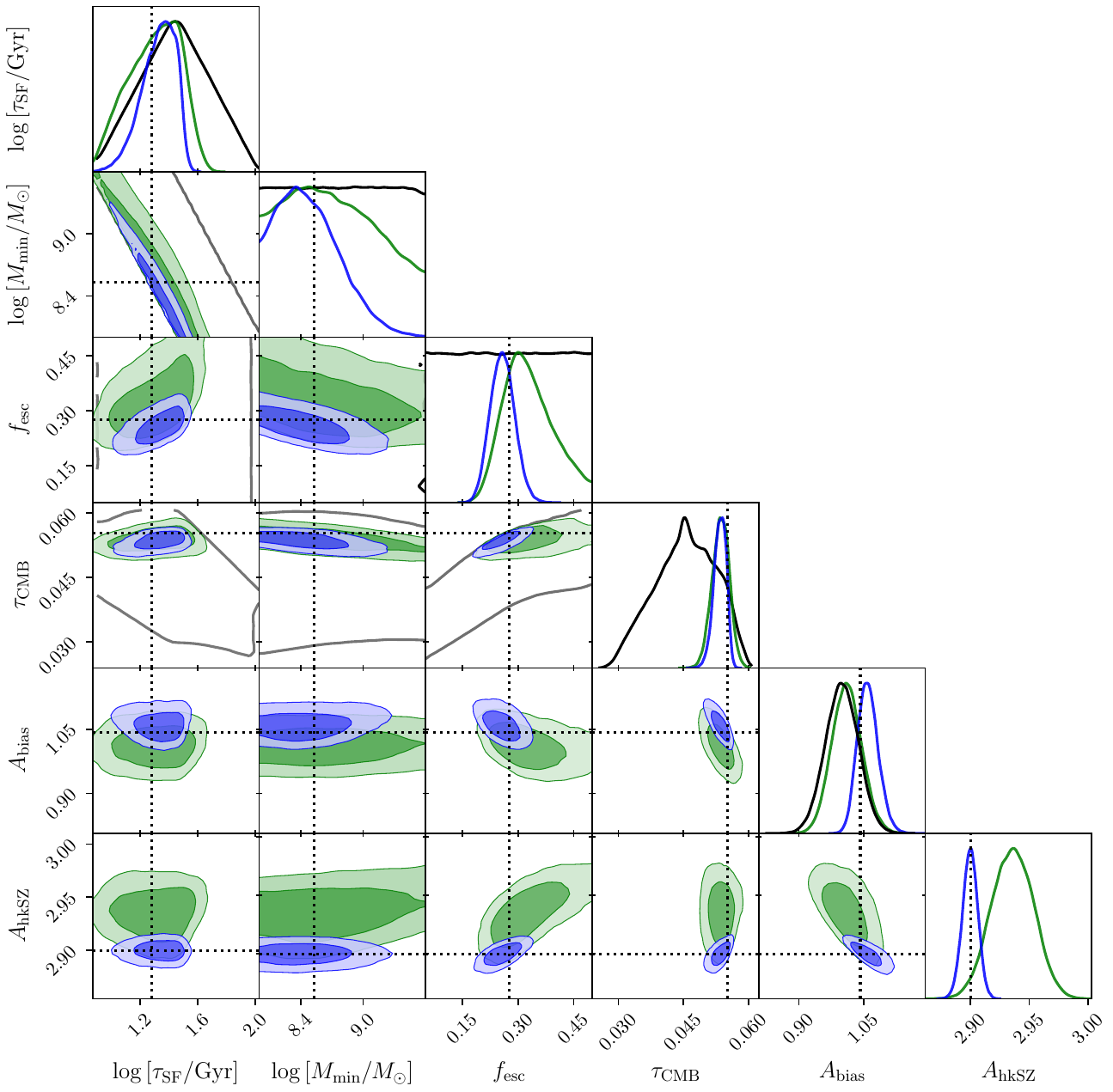}
\caption{Corner plot of MCMC-based forecasts on the parameters described in Sec.~\ref{subsec:forecast} for fitting mock data for a SO-LAT like (green) and Stage\,IV+ (blue) pair of surveys, as described in Sec.~\ref{subsec:surveys}. We show the fitted parameters along with the derived parameter, $\tau_{\mathrm{CMB}}$. For the three fitted astrophysical parameters, the derived parameter $\tau_{\mathrm{CMB}}$, and the Gaussian prior on $A_{\mathrm{bias}}$, the prior distribution is underlaid in grey for 2D and black for 1D.}
\label{fig:forecasts}
\end{figure*}

\subsubsection{Case study} \label{subsec:case}
Here we present parameter constraints obtained fitting mock data derived from a single parameter set, following the methodology outlined in Sec.~\ref{sec:methods}, assuming observations with two telescopes (Table~\ref{table:telescopes} and associated errors in Fig.~\ref{fig:surveys}). This parameter set was chosen because it leads to a Thomson optical depth compatible with \textit{Planck} results \citep{BATMAN!}.
The recovered parameter values are reported in Table~\ref{tab:bestfits} with associated marginalised posterior distributions shown in Fig.\,\ref{fig:forecasts}. \\

Crucially, for both survey configurations considered, the posterior 68\% credible region is narrower than the prior range, indicating that kSZ power spectrum measurements can constrain the input astrophysical parameters.
After prior flattening, for a Stage IV+ survey we forecast measurements of $\log{\tau_{\mathrm{SF}}/\mathrm{Gyr}} = 1.28 \pm 0.15$ for the logarithm of the gas conversion timescale, $\log{M_{\mathrm{min}}/M_{\odot}} = 8.48\pm0.31$ for the logarithm of the minimum halo mass, and $f_{\mathrm{esc}} = 0.26\pm0.03$ for the ionising escape fraction. For a less sensitive instrument such as SO-LAT, measurements are achieved with error bars 1.5 to 2.3 times larger, depending on the parameter: we get $\log{\tau_{\mathrm{SF}}/\mathrm{Gyr}} = 1.18\pm0.24$ timescale, $\log{M_{\mathrm{min}}/M_{\odot}} = 8.72\pm0.50$, and $f_{\mathrm{esc}} = 0.33\pm0.07$. The normalised bias is $\vert b^* \vert < 1$ for all astrophysical parameters across all surveys, indicating that the true value lies within the 68\% credible interval. As stated, we also reconstruct the posterior distribution of the CMB optical depth, $\tau_{\mathrm{CMB}}$, which we recover within $1\sigma$ ($|b^*| \gtrsim 1$) for both surveys considered, with an estimated uncertainty for SO-LAT of $\sim 0.0023$, roughly a factor of three improvement over current large-scale measurements \citep{BATMAN!}.

All 1D marginalised posteriors are unimodal and centred near---albeit not exactly on---the true value. Additionally, the distributions are non-Gaussian, with several parameters (notably $\log{\tau_{\mathrm{SF}}}$ and $\log{M_{\mathrm{min}}}$) having skewed posteriors. 
Between these two parameters there is a clear degeneracy, as expected from their competing effects on star formation (Sec.~\ref{subsec:params}). Despite this degeneracy, the 68\% credible interval of the posterior distribution is significantly more compact than the region contained in the coupled $\tau_{\mathrm{SF}}$/$M_{\text{min}}$ prior strip, indicating that the kSZ signal can still provide individual constraints on these two parameters. \\

For the Stage IV+ survey, the amplitude of the homogeneous kSZ power spectrum is measured with good precision, within $0.5\%$. The inclusion of the nuisance parameter, $A_{\mathrm{bias}}$, effectively mitigates the mismatch in the pkSZ amplitude induced by the emulator, which aids in the recovery of the hkSZ amplitude. In contrast, for the SO-LAT like experiment the recovered $A_\mathrm{hkSZ}$ is biased high ($b^* = 2$), whilst $A_{\mathrm{bias}}$ is biased low. A follow up run where the Stage IV+ analysis was run without data at multipoles $\ell > 5000$ shows the same qualitative behaviour, indicating that higher multipoles, where the contribution from the homogeneous signal is larger than that of the patchy, are important for fixing $A_{\mathrm{hkSZ}}$. Indeed, at low multipoles, both surveys are emulator error dominated, with only SO-LAT errors becoming dominated eventually by experimental uncertainty on scales $\ell\geq4000$. However, this amplitude mismatch for SO-LAT does not lead to extra bias in the recovered astrophysical parameters.

The relative shapes of the homogeneous and patchy components of the total kSZ can have an impact on the parameter recovery. The more similar the shapes, the more difficult it is to separate the two kSZ contributions. Therefore, we perform a second forecast, this time assuming a different template for the homogeneous kSZ \citep{Shaw_2012}, which exhibits less power at high multipoles than our fiducial template, such that its shape is closer to the patchy spectrum. We find little impact, especially on the astrophysical parameter posteriors, in switching hkSZ templates. Only the $A_{\mathrm{hkSZ}}$ posterior exhibits slightly wider contours for the case where the template shape is more similar. We also note that in this work we ignore the impact of the astrophysical parameters on the amplitude of the hkSZ amplitude, which in practice, depends on the ending redshift of reionisation (with earlier ends boosting the hkSZ amplitude). To investigate the impact of the amplitude on the resultant fits, several runs were conducted while varying the amplitude within the predicted range and in each case, the hkSZ amplitude can be recovered, however future work should consider a fully correlated hkSZ + pkSZ model.\\

\subsubsection{Statistical analysis} \label{subsec:stats}

\begin{figure*}
\centering
\includegraphics[width=0.8\textwidth]{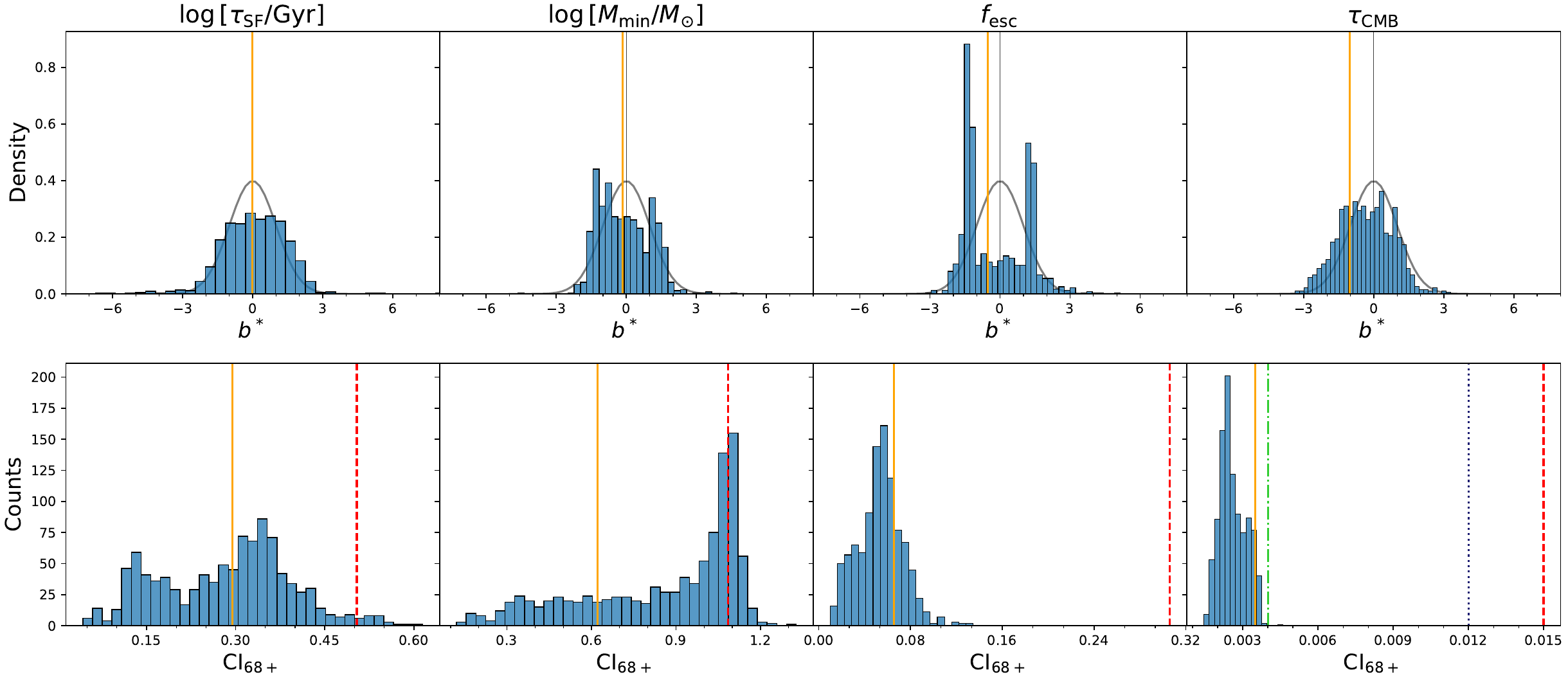}
\caption{Statistical analysis of the performance of our inference pipeline after prior flattening for 10\,000 inference runs. \textit{Top panel:} Histograms of the normalised bias, $b^*$ (Eq.\,\ref{eq:normedbias}), underlaid with a standard Gaussian (gray line). \textit{Bottom panel:} Histograms of the width of the conservative 68\% CI ($\mathrm{CI_{68+}}$) of each 1D marginalised posterior, normalised by the 95\% CI of of the prior distribution. The values corresponding to the fit of our fiducial parameter set are overlaid (gold line, Sec.~\ref{subsec:case}) along with the \textit{Planck} $1\sigma$ constraints (dotted indigo), and the cosmic variance limit from large scale surveys (green dash dotted) in the bottom right plot. The 68\% CI derived from the prior distribution (without prior flattening) is also shown (dashed red).}
\label{fig:biases}
\end{figure*}

While promising, we cannot generalise the constraints from our above case study to the whole data set. Since the emulator performance varies between simulations (Fig.\,\ref{fig:kemu}) and because of the shape of the coupled $\tau_{\mathrm{SF}}$/$M_{\text{min}}$ prior, we expect the accuracy of our inference to vary. In order to characterise the overall accuracy of our parameter recovery, we analyse the results of 1000 inference runs performed for random sets of `true' parameter values, assuming Stage IV+ observations, in an approach somewhat analogous to simulation-based calibration \citep[SBC, ][]{2018arXiv180406788T, MeriotSemelin_2025}.
We prior flatten the resulting samples, as done for our case study.


In the top row of Fig.~\ref{fig:biases}, we show the distribution of the 1000 normalised biases for each parameter. If our emulator accuracy was limited by simple Gaussian errors (e.g., propagated from cosmic variance), we would recover a mean-zero normal distribution with a standard deviation $\sigma \sim 1$ (dotted line). Instead, our distributions are less peaked, which indicates that our inference accuracy is low ($|b^*| > 1$) more often than expected if the emulator error was normally distributed. For the three fitted astrophysical parameters, $\log{\tau_{\mathrm{SF}}}$, $\log{M_{\mathrm{min}}}$, and $f_{\mathrm{esc}}$, the best fit values are biased ($|b^*| \geq 1$) 47\%, 46\%, and 77\% of the time, respectively.

In addition to accuracy, it is also necessary to assess how much the precision of our constraints varies across simulations. In the bottom row of Fig.\,\ref{fig:biases}, we show histograms of the widths of our conservative  credible interval $\mathrm{{CI}_{68+}}$. The posterior distributions width varies over our runs, following the location of the parameter values within the prior and the emulator accuracy across the parameter space.
Overall, the mock data do provide information about the astrophysical parameters: 
The gas conversion timescale and the ionising escape fraction are consistently recovered with credible intervals smaller than the prior range, with a mean reduction of a factor approximately 2 and 4 in $\mathrm{{CI}_{68+}}$, respectively. However, the size of the credible interval for the former has a larger dispersion than the latter; all samples for $f_{\mathrm{esc}}$ show a factor of two reduction in parameter space compared to the $\log{\tau_{\mathrm{SF}}}$, which is prior-dominated in a few cases, though by less than 3\%.

Unlike our case study, where the minimum halo mass is recovered with good precision, we observe that for many parameter sets, the data provides little additional information on $M_{\mathrm{min}}$ beyond what is already contained in the prior.  
This could be because of the coupling in parameter space between the gas conversion timescale and the minimum halo mass: The data is not able to break the degeneracy between the parameters, such that only one of the two is constrained (see Fig~\ref{fig:forecasts}). 

In the right-most column of Fig.\,\ref{fig:biases} we also show the results for the derived $\tau_{\mathrm{CMB}}$. The distribution of the normalised biases is relatively symmetric, mean-zero centred, with 60\% of simulations in our statistical sample having $|b^*| \leq 1$. The mean fractional reduction of the allowed parameter range is $\mathrm{CI}_{68+} / \mathrm{CI}_{\mathrm{priors, 68}} = 0.16 \pm 0.03$, corresponding to an average uncertainty of $\sigma_{\tau_\mathrm{CMB}} = 0.0012 \pm 0.0002$. Notably, this indicates that measurements of the CMB reionisation optical depth can be improved beyond the cosmic-variance limit of low-$\ell$ measurements \citep{Litebird}. 


\section{Discussion} \label{sec:discuss}

While our results demonstrate the potential of constraining astrophysics through the kSZ power spectrum, this analysis is limited by unavoidable features of our framework, that is the emulator accuracy and the distorting effect the coupled $\tau_{\mathrm{SF}}$/$M_{\text{min}}$ prior imprints on the posterior distributions of our parameters. 
First, the emulator accuracy is hindered by the size and spread of the training dataset. 
For example, the poor recovery of the escape fraction, whose $b^*$ distribution has a bimodal structure, is likely due to the sparsity of the training set. Only three values of $f_\mathrm{esc}$ are considered, compared to 42 for $\log{\tau_{\mathrm{SF}}}$. 
This is due to both the original scope of the simulations--- namely the study of the 21\,cm signal during Cosmic Dawn, which is strongly impacted by the two X-ray parameters, in contrast to the kSZ--- and to the gridded spacing of the input parameters. A finer, more comprehensive coverage of the parameter space (for example, Latin hypercube sampling) would improve emulator accuracy. For the particular case of 21\,cm and reionisation-era kSZ studies, we note that even a modest increase in coverage of the ionising parameter(s) would improve the complementarity of such simulations. 

Second, the amount of bias we observe in our parameter recovery (Fig.~\ref{fig:biases}) is also affected by the prior range, and in particular the shape of the $\tau_\mathrm{SF}-M_\mathrm{min}$ prior, which artificially distorts the posterior distributions. True values that lie near any prior edge tend to result in biased inference, due to the inability to sample beyond the border. 
Again, this effect is strongest for the ionising escape fraction, wherein the values that lie on the edge of the parameter range are biased 100\% of the time compared to the central value ($f_\mathrm{esc}=0.275$, also in the case study) which is biased 49.5\% of the time. Thus, relaxing this strict prior via a wider range of input parameter values would soften this effect. Future analyses could also consider moving to Simulation-Based Inference (SBI) as investigated in \cite{MeriotSemelin_2025}.


Further progress could be made by improving the kSZ modelling itself. The formalism described in Sec.~\ref{subsec:Gorce} requires access to both the electron overdensity power spectrum for all Fourier modes, $k$, and the full ionisation history, $x_e(z)$. However, due to the finite simulation volume, the \textsc{LoReLi~II} power spectra are restricted to a limited $k$-range, accessing scales $k>0.031 ~\mathrm{Mpc}^{-1}$. Although we have found these missing modes to not significantly impact our reconstructed patchy kSZ amplitude (less than $2\%$, see appendix~\ref{appendix:cuts}), the shape of the $P_{ee}$ spectra on large scales is unknown and the missing large scales could have a stronger impact than measured here \citep{Park}.
Additionally, the patchy kSZ signal depends on the ionisation history up to full hydrogen ionisation, however due to computational constraints the final few percents are not simulated in the \textsc{LoReLi~II} database. In appendix~\ref{appendix:cuts}, we find that missing the final 3\% of the ionisation history can impact the patchy kSZ amplitude at the $5\%$ level. Addressing both these restrictions, for example by increasing the simulated cosmological volume, or extending the simulation up to a higher ionisation fraction, would improve the accuracy of the kSZ reconstruction.
Further modelling refinements could involve moving beyond some of the approximations in Sec.~\ref{subsec:Gorce}, that is measuring the velocity power spectrum instead of relying on a linear theory approximation (see \citet{GorceIlic_2020} for further details). Another approximation is the use of coeval simulation cubes and the Limber approximation: computing the kSZ signal by ray tracing in simulated light cones would better capture the evolution of the signal, particularly by retaining long wavelength mode correlations.
Finally, we are able to derive meaningful constraints from an emulator trained on a sub-optimal training set, showing that the \textsc{LoReLi~II} database, which was not designed for kSZ studies, can serve as a complementary probe to other analyses. Thus, our work highlights the synergistic potential of repurposing existing simulations for use beyond their original science case. Further joint optimisation for studies of this epoch could maximise the utility of computationally expensive simulation suites and help reduce the carbon footprint of astrophysics research \citep{carbon}.

Altogether, the modelling landscape for the kSZ power spectrum remains relatively broad. It is therefore useful to compare the results presented here with ones obtained with alternative models. One such example is the semi-analytic model proposed by \citet{ourfrenemies} [J24 hereafter], where the kSZ spectrum is derived using simulations of the ionisation field across reionisation, obtained with the \texttt{SCRIPT} simulation code \citep{Choudhury_2018}. 
Since the astrophysical parameterisation differs between \texttt{SCRIPT} and \textsc{LoReLi~II}, a broad direct comparison is not possible. However, both approaches involve running inference on mock data of the kSZ power spectrum in order to fit a set of astrophysical parameters. In doing so, J24 also arrive at the conclusion that the kSZ power spectrum can meaningfully constrain astrophysical properties. Some differences with our work should, however, be noted: their observational errors, obtained with Cross-internal linear combination \citep[ILC,][]{raghunathan2023crossinternallinearcombinationapproach}, are larger, and their analysis covers a more compact multipole range, $\ell=[2500,5000]$.
One point of comparison is the recovery of the Thomson optical depth, $\tau_{\mathrm{CMB}}$. The forecasted constraints in J24 exhibit less bias, with a normalised bias of roughly $\sim 0.8$  (compared to $\sim 1.1$ for our result) for their SO-Goal and S4-Wide experimental setups. 
Their tightest constraint on $\tau_{\mathrm{CMB}}$ (with the inclusion of Planck data) comes from assuming a S4-Wide survey, with a forecasted error of $\sim 0.0053$. This is larger than our forecasted error for a Stage IV+ survey of $\sim 0.0018$, however we note that our foreground assumptions are more optimistic. Additionally, we forecast measurements of the reionisation duration (defined as $\Delta z = z_{\langle x_\text{HII} \rangle = 25} - z_{\langle x_\text{HII} \rangle = 75}$) with uncertainty $\sim 0.0275$ for SO-LAT and $\sim 0.0190$ for a Stage IV+ telescope. This is much more precise, about an order of magnitude, than the constraints presented in J24 and may result from differences in the modelling approach. J24 are also unable to constrain both their minimum halo mass parameters. 
Overall we find good qualitative agreement between the two works. 

\section{Conclusions} \label{sec:fin}

In this work, we use the \textsc{LoReLi~II} database of cosmological simulations to reconstruct the patchy kSZ power spectrum as a function of five input astrophysical parameters,  $\vec{\theta} = \{f_X, \mathrm{r_{\mathrm{H/S}}}, \tau_{\mathrm{SF}}, M_{\mathrm{min}}, f_{\mathrm{esc}}\}$. We then use our reconstructed kSZ dataset, composed of 6796 simulations, to investigate the dependence of the kSZ signal on the astrophysics of the sources responsible for reionisation (Fig.~\ref{fig:kSZ}), and additionally the potential of current and future CMB ground-based observatories to constrain said astrophysics. We do this by constructing a neural-network based emulator (Fig.~\ref{fig:kemu}) in order to perform a Bayesian analysis, fitting the three parameters with dominant impact, $\vec{\theta}_{\mathrm{fit}} = \{ \log{\tau_{\mathrm{SF}}}, \log{M_{\mathrm{min}}}, f_{\mathrm{esc}}\}$.

In Sec.~\ref{subsec:case}, for a single case study, we find that we can constrain all fitted parameters, and the additional derived parameter the reionisation optical depth $\tau_{\mathrm{CMB}}$, to within credible intervals smaller than those imposed merely by the prior distribution itself (Table\,\ref{tab:bestfits} and Fig.~\ref{fig:forecasts}). To generalise these results, in Sec.~\ref{subsec:stats} we run 1000 forecasts for 1000 different sets of `true' parameter values. We find that the mock data can consistently constrain both the gas conversion timescale, $\log{\tau_{\mathrm{SF}}}$, and ionising escape fraction, $f_{\text{esc}}$, 97\% and 100\% of the time, respectively, 
but are only minimally sensitive to the minimum halo mass, $\log{M_{\mathrm{min}}}$, which is prior-dominated in about 30\% of the cases. 
For the ionising escape fraction --- our most dominant parameter --- we forecast an average relative uncertainty of 14\%. Additionally, we are able to constrain the reionisation optical depth $\tau_{\mathrm{CMB}}$ with a modest average bias of $b^* = 0.43$ (Eq.~\ref{eq:normedbias}). Across the 1000 runs, the optical depth is measured with an mean uncertainty of $\sigma_{\tau_{CMB}} \sim 0.0012$, below the cosmic-variance limit ($\sigma_{\text{cv}} = 0.002$), and with a maximum uncertainty of $\sigma_{\tau_{CMB}} \sim 0.0023$. 
Such a measurement would represent a fivefold reduction over current constraints from CMB large-scale analyses.

While we demonstrate that measurements of the kSZ power spectrum by the Simons Observatory can yield meaningful astrophysical constraints, the precision of the results improves with Stage IV+ surveys, and a future survey such as proposed by \cite{CMB-HD} has the potential to meaningfully improve constraints. 
However, our results are limited by the accuracy of our emulator. 
In Sec.~\ref{sec:discuss}, we discuss both these limitations and possible avenues for improvement, which mostly require finer and wider parameter sampling in the training set. 
Even given current observational and modelling limitations, we show that the kSZ power spectrum is a sensitive probe of astrophysics, and can provide meaningful information about the ionising sources in the early Universe. 
%


\begin{acknowledgements}
The author(s) acknowledge(s) the support of the French Agence Nationale de la Recherche (ANR), under grant ANR-22-CE31-0010 (project BATMAN). This work was supported by the ``action thématique" Cosmology-Galaxies (ATCG) of the CNRS/INSU PN Astro and ANR PIA funding ANR-20-IDEES-0002. We also acknowledge the use of many open-source libraries, including \texttt{Numpy}\citep{numpy}, \texttt{Matplotlib} \citep{matplotlib},  \texttt{Astropy} \citep{astropy:2013, astropy:2018, astropy:2022}, \texttt{Scipy} \citep{scipy}, \texttt{Powerbox} \citep{powerbox}, and \texttt{anesthetic} \citep{anesthetic}. All code used in this work is available for use and inspection.
While used for short form wording suggestions, no significant portion of this work was processed using AI tools --- all emdashes were the choice of the authors.
\end{acknowledgements}

\bibliographystyle{aa}
\bibliography{biblio}

\appendix{}
\section{Impact of astrophysical parameters on the electron overdensity power spectrum} \label{appendix:astroparams}
\begin{figure*}
\centering
\includegraphics[width=.8\textwidth]{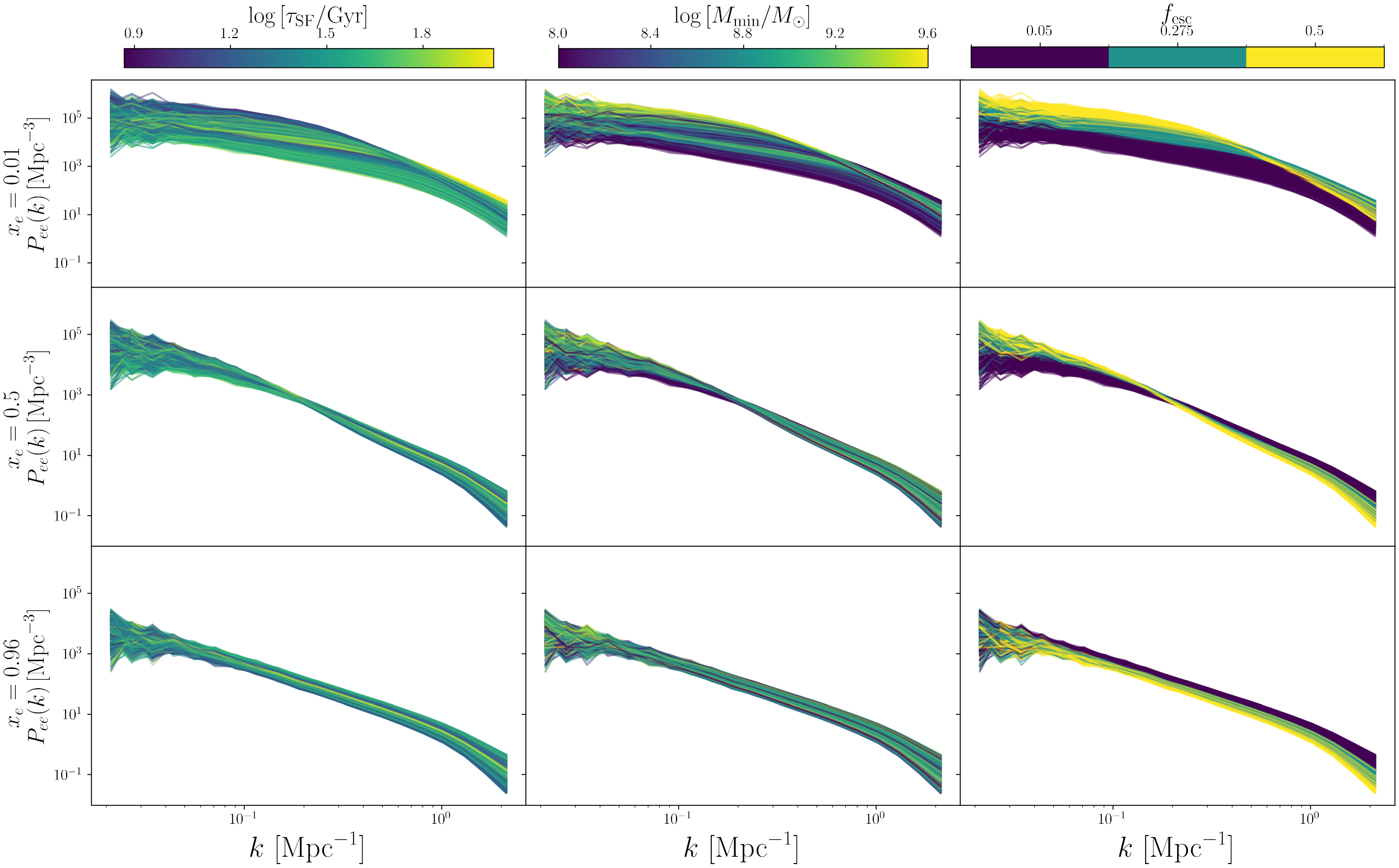}
\caption{Electron overdensity power spectra derived from the \textsc{LoReLi~II} database, shown at the beginning (top row), midpoint (middle row), and end (bottom row) of reionisation. The spectra are coloured by the value of three astrophysical parameters varied in the dataset in each column. The clearest trend is with the ionising escape fraction (third column), with some observed trends due to the minimum halo mass (second column).}
\label{fig:LoReLi_full}
\end{figure*}

Here we show the redshift evolution of the impact of astrophysical parameters on the electron overdensity power spectra, $P_{ee}(k,z)$, one of the two data products derived from the \textsc{LoReLi~II} database. In Fig.\,\ref{fig:LoReLi_full}, we show $P_{ee}$ at three hydrogen ionisation fractions, corresponding to three critical regimes within the timeline of the EOR; the beginning (1\%), midpoint (50\%), and end (96\%) of reionisation. The evolution of the electron overdensity power spectra depends most clearly on the ionising escape fraction (third column), but is also correlated with the minimum halo mass (second column).

\section{Impact of the coupled $\tau_{\mathrm{SF}}$/$M_{\text{min}}$ prior} \label{appendix:2dprior}
\begin{figure}
\centering
\includegraphics[width=\columnwidth]{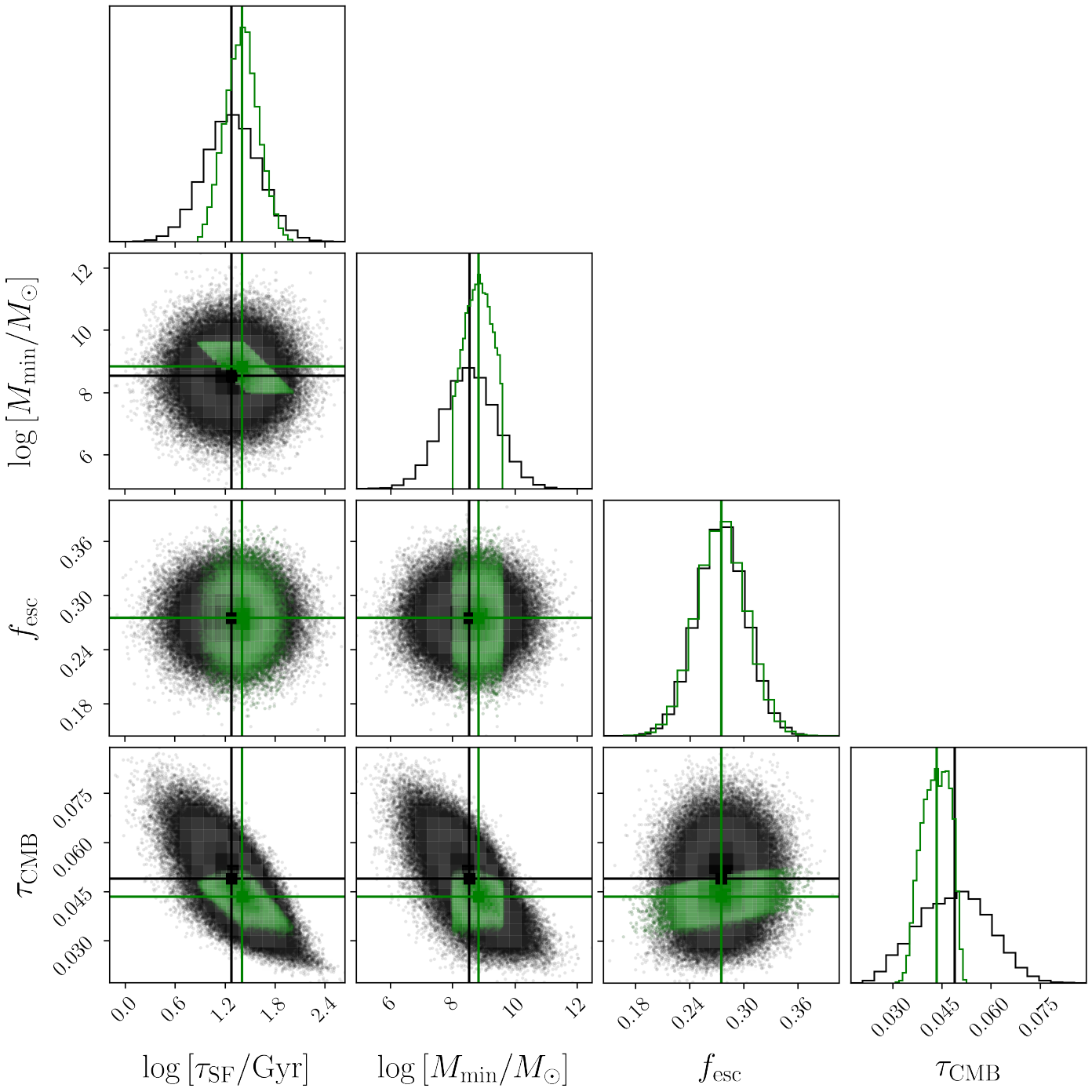}
\caption{Impact of the coupled $\tau_{\mathrm{SF}}$/$M_{\text{min}}$ prior on a normally distributed sample of parameter values. In black, we show samples drawn from a Gaussian distribution centred on the parameter values used for the first mock dataset (Sec.~\ref{subsec:case}). Green samples are obtained after filtering the Gaussian distributions with our prior. The mean value for a set of values is shown as a vertical line in the corresponding colour. }
\label{fig:2dprior}
\end{figure}

 Because we observe that the coupled $\tau_{\mathrm{SF}}$/$M_{\text{min}}$ prior restricts the parameter range that can be explored by the MCMC sampler, we investigate its impact on our posteriors. While difficult to quantify, we observe some general trends across the parameter space:
 \begin{itemize}
    \item The width of our contours is artificially reduced because of a sharp truncations in the histograms at the prior edge.
    \item The mean values are biased from said truncation.
    \item The impact varies across the parameter space, as the posteriors of parameters that lie close to a prior bound will be more strongly affected. 
\end{itemize}
In Fig.\,\ref{fig:2dprior} we show the impact of implementing the coupled $\tau_{\mathrm{SF}}$/$M_{\text{min}}$ prior on normal distributions centred on the true values of our fiducial simulation. The impact on the posterior is clear --- the tails of the Gaussian distributions are cut off, which compresses the distributions and shifts their mean.

However, it is difficult to extrapolate these results to our posterior distributions (Fig.~\ref{fig:forecasts}), which are clearly non-Gaussian. When compared with our fiducial run, the Gaussian samples are simply truncated by the prior, whereas the MCMC sampler expands out from the central, high probability region. This is enabled by the degeneracy between the gas conversion timescale and the minimum halo mass, which flattens the likelihood function, making it easier for the walkers to traverse the parameter space without being overly penalised. This widens the contours for some parameters, notably the minimum halo mass $\log{M_{\text{min}}}$ . 

\section{The effects of X-ray parameters on the kSZ signal} \label{appendix:Xrays}
As stated in Sec.~\ref{subsec:params}, we do not expect the X-ray parameters to have a large impact on the kSZ signal, as X-rays are too energetic to efficiently ionise the IGM and are instead primarily responsible for heating of gas within the IGM. 
However, in certain cases, we see a stronger than expected impact of the X-ray parameters on the kSZ power spectrum, mostly due to changes in the ionisation history. One potential source of this enhanced correlation is in the implementation of the ionising escape fraction parameters. Within the \textsc{Licorice} code, there are two separate parameters which dictate the amount of ionising flux emitted by haloes into the IGM, namely the pre- and post-threshold escape fractions. If a halo is in a region where the density of ionised hydrogen is below the critical threshold $x_\ion{H}{II} = 3\%$, then the pre-threshold value of $f_{\text{esc, pre}}=.003$ is used. However, once the ionised hydrogen fraction rises above this threshold, the escape fraction switches to the input value of $f_{\text{esc}}$. While a small effect, local heating of the IGM can increase the ionising efficiency enough to bump $f_{\text{esc, pre}} \to f_{\text{esc}} $ earlier for a high X-ray scenario than a low one, leading to a more aggressive transition than what would occur naturally. This model may explain the spurious dependence on the X-ray parameters observed in some cases. However, because this dependence is likely driven by the implementation within the simulation we choose to ignore it in any further analysis.

\section{Effects of cuts on the kSZ spectra} \label{appendix:cuts}
\begin{figure}
\centering
\includegraphics[width=\columnwidth]{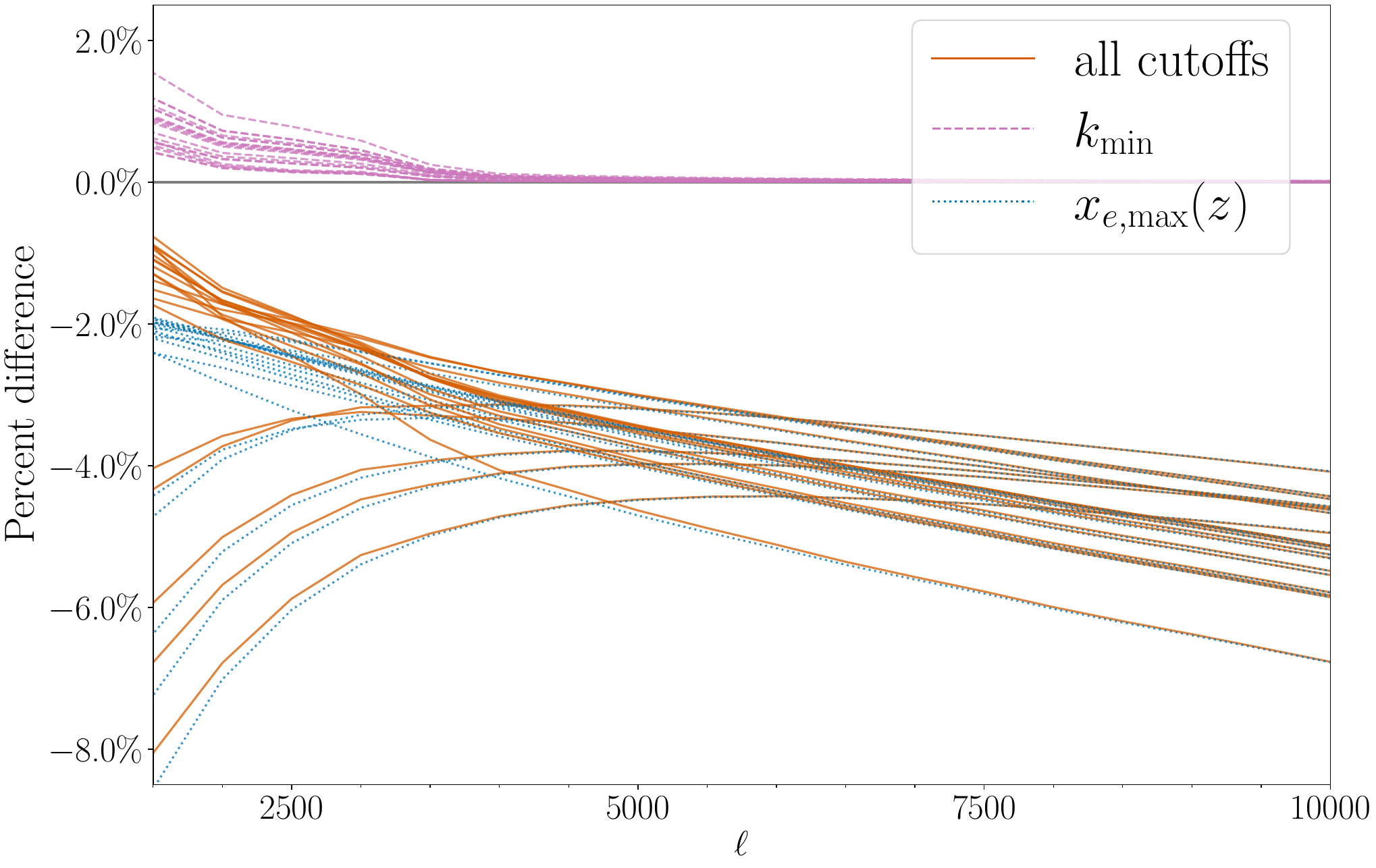}
\caption{Changes to the signal power for a sample of kSZ power spectra due to truncations of the data from the loss of large scale Fourier modes (violet), the missing end of ionisation history (blue), and the total impact on the signal (orange). Results are shown for a random draw of ten parameter sets.}
\label{fig:cuts}
\end{figure}

The limited physical size of the \textsc{LoReLi~II} simulations, and the restricted redshift range they cover, affect the kSZ power spectrum we build from their electron power spectra. Here, we investigate the consequences of these limitations by using the $P_{ee}$ parameterisation proposed in \cite{GorceIlic_2020}, defined for any redshift and Fourier scale. This means we can apply the same cuts to this model as the ones imposed by our dataset, and compare the resulting kSZ signal against the one obtained without cuts. The resulting ratios are shown in Fig.~\ref{fig:cuts} for a random draw of 10 parameter sets across the database.

First, we look at the impact of restricting the kSZ integral (Eq.~\ref{eq:Cell}) to redshifts when $x_\mathrm{HII}<0.97$ (as many simulations do not reach higher ionisation fractions), in blue in the figure. 
This early truncation results in an average $\sim5\%$ loss in power across the multipole range and is the main source of signal loss. 

Second, we take into account the finite size of the simulated volume by removing the contribution of wave-modes larger than the simulations ($k<k_\mathrm{min}$) from the kSZ integral in Eq.~\eqref{eq:Cell}. Removing their contribution increases the reconstructed power at low multipoles. This is because the $P_{ee}$ amplitude decreases, such that the negative cross-term of Eq.~\eqref{eq:PBe} becomes dominant. 
However, this effect is small ($\sim 2\%$) and confined to low multipoles.




%
%
\end{document}